\documentclass[sigconf]{aamas} 

\usepackage{amsfonts}       
\usepackage{amsmath}
\usepackage{pifont}
\usepackage{dsfont}
\usepackage{algorithm}
\usepackage{algorithmic}
\usepackage{ragged2e}

\usepackage{tikz}
\usepackage{tikz-3dplot}
\usepackage{pgfplots}
\pgfplotsset{compat=1.16}
\usetikzlibrary{patterns.meta,positioning,arrows.meta,calc,shapes.geometric, matrix, fit,decorations.pathreplacing}

\newcommand{\cmark}{\ding{51}}%
\newcommand{\xmark}{\ding{56}}%

\newcommand{\set}[1]{\mathcal{#1}}
\newcommand{\setS}{\set{S}}
\newcommand{\setSig}{\Sigma}
\newcommand{\setR}{\set{R}}
\newcommand{\setA}{\set{A}}
\newcommand{\setAs}{\setA(s)}
\newcommand{\setAlph}{\mathrm{A}}
\newcommand{\setAsig}{\setAlph(\sigma)}
\newcommand{\setU}{\set{U}}
\newcommand{\setUs}{\setU_s}

\newcommand{\setN}{\mathbb{N}}
\newcommand{\setNs}{\mathbb{N}^*}
\newcommand{\setNsn}[1]{\mathbb{N}^*_{#1}}
\newcommand{\setNn}[1]{\mathbb{N}_{#1}}
\newcommand{\ind}[1]{\mathds{1}\left[#1\right]}

\newcommand{\bigO}[1]{\set{O}\left(#1\right)}
\DeclareMathOperator*{\argmax}{arg\,max}
\DeclareMathOperator*{\esp}{\mathbb{E}}
\newcommand{\pluseq}{\mathrel{+}=}

\usepackage{balance} 

\setcopyright{ifaamas}
\acmConference[AAMAS '26]{Proc.\@ of the 25th International Conference
on Autonomous Agents and Multiagent Systems (AAMAS 2026)}{May 25 -- 29, 2026}
{Paphos, Cyprus}{C.~Amato, L.~Dennis, V.~Mascardi, J.~Thangarajah (eds.)}
\copyrightyear{2026}
\acmYear{2026}
\acmDOI{}
\acmPrice{}
\acmISBN{}

\acmSubmissionID{211}

\title[pMDPs, Optimizing LRA in Blockchains]{Pseudo-MDPs: A Novel Framework for Efficiently Optimizing Last Revealer Seed Manipulations in Blockchains}

\author{Maxime Reynouard}
\affiliation{
  \institution{Nomadic Labs \& Université Paris Dauphine - PSL}
  \city{Paris}
  \country{France}}
\email{maximereynouard@gmail.com}

\begin{abstract}

This study tackles the computational challenges of solving large Markov Decision Processes (MDPs) for a \textit{restricted} class of problems with reversed decision flows, a structure that arises in repeated two-stage stochastic control and in applications such as taxi dispatch and the Last Revealer Attack (LRA) on Ethereum’s RANDAO. We focus on the LRA, which undermines fairness in certain Proof-of-Stake (PoS) blockchains such as Ethereum (\$400B market capitalization). Standard modeling of this problem as an MDP would lead to a state space with a size of the order of $10^{10^{10}}$ and stochastic inputs as large as  $5\,\mathrm{GB}$, making it intractable. We introduce the notion of pseudo-MDP (pMDP), a framework that naturally models such problems and propose two distinct reductions to standard MDPs.
The duality of the reductions provides a novel perspective, enabling significant improvements in value iteration algorithms, including  a complexity compression close to logarithmic for the dynamic programming implementation and a more efficient Monte Carlo variant. In the case of the LRA which size is parameterized by $\kappa$ (in Ethereum's case $ \kappa=32 $), we reduce the computational complexity per iteration from $\bigO{ 2^\kappa\, \kappa^{2^{\kappa +2}} }$ to $ \bigO{ \kappa^4} $.
This perspective also simplifies policy extraction, making the approach well-suited for resource-constrained agents, once again for the LRA, we lower the worst-case complexity from: $\bigO{ 2^\kappa\, \kappa^{2^{\kappa +1}} }$ to $\bigO{ 2^\kappa }$. The application case of the LRA demonstrates the capability of the pMDP framework to solve large scale problems effectively.
Our approach also provides the usual benefits from dynamic programming solutions: exponentially fast convergence toward the optimal solution is \textbf{guaranteed}. The efficiency we achieve comes at the cost of the framework's applicability, which is why we also provide an extension of some results to a broader class of MDPs than that of reduced pMDPs.

\end{abstract}


\begin{document}


\pagestyle{fancy}
\fancyhead{}


\maketitle 


\section{Introduction}

Blockchains enable smart contracts and decentralized applications, transforming sectors like finance, gaming, and supply chains~\cite{SurveyBlockchainApp,BlockchainAppSupplyChain}. Unlike Proof of Work (PoW)~\cite{nakamoto2009bitcoin} systems that consume significant energy~\cite{bitcoinConsumption,EthConsumption}, Proof of Stake (PoS) blockchains select consensus participants based on their stake. To guarantee fairness and security, PoS protocols rely on reliable randomness sources. Several prominent PoS blockchains~\cite{EthSpecs,harmony,NearConsensus} adopted the RANDAO (Random Autonomous Organization) protocol~\cite{Randao}, which generates randomness through a decentralized, verifiable two-step \emph{commit-reveal} process~\cite{CryptoFundation}: participants first commit secret values via hashes, then reveal them to collectively produce the random outcome (typically using XOR). We specifically examine Ethereum’s RANDAO due to Ethereum’s prominence and market cap in the hundreds of billions of USD.

Time-constrained commit-reveal schemes are vulnerable to the Last Revealer Attack (LRA)~\cite{Randaomanip}, where the final participant influences the outcome by choosing whether or not to reveal their secret, enabling selection between the intended result or a fallback value. This undermines the fairness and unpredictability of randomness generation. Although widely recognized in cryptographic research, pinpointing the origin of this vulnerability is challenging. Several mitigation strategies exist, such as incentive design, Threshold Schemes~\cite{RandaoThreshold}, or Verifiable Delay Functions~\cite{VDF,VDFmini}, but exploitation remains a credible threat~\cite{Randaomanip}. Until recently, quantifying the advantage from LRA was difficult~\cite{OptimalRandaoManip}. The existing solution, while innovative, applies only specifically to RANDAO's LRA and is not easily generalized. Our approach, in addition to being general, provides a slight algorithmic improvement, reducing complexity from $\bigO{\kappa^5}$ to $\bigO{\kappa^4}$.

Markov Decision Processes (MDPs)~\cite{mdp, mdpBellman} are natural tools to model and optimize this attack, but the combinatorial complexity makes their use challenging — Ethereum’s case would require a prohibitive number of states ($\in\bigO{\kappa^{2^{\kappa+1}}}$). Factored MDPs~\cite{factoredMDP} could improve representation, though they still involve products of $2^{\kappa}$ sets and do not simplify solving the MDP. Standard dynamic programming methods solve small MDPs efficiently but struggle at scale. Heuristic or approximate methods like Monte Carlo Tree Search or neural-network-based function approximation can manage larger state spaces but cannot effectively exploit structure due to the Cryptographically Secure PseudoRandom Number Generators (CSPRNG)~\cite{introCrypto}, which guarantee high entropy. Additionally, such approaches only yield approximate solutions without strong theoretical guarantees.

The difficulty of modeling RANDAO's LRA using MDPs stems from an inversion in the natural flow of decisions. Traditionally, MDPs assume an agent selects an action that leads probabilistically to a new state, the flow follows:

\centering\noindent "state" $\rightarrow$ "decision" $\rightarrow$ "stochastic process" $\rightarrow$ "new state".	

\justifying\noindent However, the LRA scenario reverses this logic: when the attacker controls the last $k$ block of an epoch, $2^k$ random seed are randomly generated, the attacker chooses one which yields a reward and gives him control over $2^{k'}$ blocks for the next epoch. This problem is more intuitively modeled by the pseudo-MDP (pMDP) framework which mimics the reversed flow:

\noindent\centering"state"$\rightarrow$ "stochastic process" $\rightarrow$ "decision"  $\rightarrow$ "new state".

\justifying\noindent From a given state $\sigma$, a stochastic process first generates the list of possible actions $\setA$, and then the agent selects one $a\in\setA$ which will yield a reward $R$ and lead him to a new state $\sigma'$. Note that the list of actions could be seen as another type of state: an \textit{ex-post state} $s$; by opposition to the \textit{ex-ante state} $\sigma$ before the stochastic process. Such a structure is vaguely similar to that of two-stage stochastic control problems~\cite{twoStage} where agents take decisions in two steps. When such problems are repeated, we can find other applications to our framework. For instance, in the case of ride selection for taxis: from a certain area $\sigma$ (ex-ante state), drivers are stochastically offered several taxi rides from their various apps and dispatch $\setA$, they select one which yields a fare $R$ and takes them to a new area $\sigma'$. It seems intuitive that there might be situations where it is worthwhile for a driver to accept a lower fare than other offers (adjusted for the ride time), if it takes them to a better area (in the sense that it will probabilistically lead to better ride proposal). This is the usual trade-off agents face in MDPs: instant reward or higher delayed reward. The distinction between \emph{ex-ante} states (before the stochastic process) and \emph{ex-post} states (after the stochastic process, matching standard MDP) also naturally arises in certain games, such as \emph{Seasons} (players choose which dice to roll, that is they choose an ex-ante state; the dice roll result form the ex-post state) and \emph{Dice Forge} (players modify dice faces thus changing future ex-ante states).

\textbf{Main contributions:} 
The main technical contributions of the paper are:
\begin{itemize}
	\item The introduction of \emph{pMDPs}, as a complementary discrete-time stochastic control process to that of MDPs~\cite{mdp}.
	They naturally model the reversed decision flow of the LRA. In doing so they fully solve the problem representation issue.
	\item Show that pMDPs can be reduced to MDPs in two different ways.  
	The reductions allows us to get all the known results of MDPs such as known solutions and their guarantees~\cite{Sutton1998}. But they do not solve the scalability issue: For instance the complexity of value iteration is in $ \bigO{ 2^\kappa\, \kappa^{2^{\kappa +2}} } $ per iteration.
	\item Lowering the complexity of value iteration in two steps, ending up in $ \bigO{\kappa^4} $ by combining the strengths of the two problem reductions.
	\item Extending those results to a larger class of MDPs where \emph{ex-ante states} act as intermediate steps in the Markov transitions.
\end{itemize}

\section{Motivating problems}
\subsection{Last Revealer Attack (LRA), case of Ethereum's RANDAO}
The Last Revealer Attack (LRA) exploits random number generation protocols, allowing the final participating agent to unfairly influence outcomes intended to be random, undermining blockchain fairness. In Ethereum's consensus protocol, \textit{Gasper}, block proposal rights are allocated every 12-second \emph{slot}. Each \emph{epoch}, composed of $\kappa=32$ slots, uses a new random seed to allocate these rights randomly, independently, and identically (IID), proportional to each agent’s stake.\footnote{Agents stake part of their Ethereum currency to participate in consensus. This staked currency is temporarily frozen as collateral to penalize misbehavior.} In RANDAO, agents first commit to a secret number (e.g., using a cryptographic hash function like SHA-256) and later reveal it when proposing blocks. These revealed numbers are combined to produce a pseudo-random seed which will be used to generate the slot allocation at the next epoch.

\tikzset{
	base/.style={
		circle,
		minimum height=0mm,
		inner sep=1pt,
		align=center,
	},
	slot/.style={
		base,
		draw=Blue, thick,
		fill = Cerulean!10,
		text = Blue,
	},
	dot/.style={
		base,
		text = Blue
	},
	check/.style ={
		base,
		minimum height =5mm,
		fill = Green,
		text = White
	},
	xmark/.style ={
		base,
		minimum height =5mm,
		fill = Red,
		text = White
	},
	qmark/.style ={
		base,
		minimum height =5mm,
		fill = ProcessBlue,
		text = White
	},
}

\newcommand*{\slot}[3]{\node[slot,shift={(#2mm,-#3mm)}] (slot#1) {slot \\[-3pt] #1};}

\newcommand{\dicenum}[1]{%
	\pgfmathparse{#1==2 || #1==4 || #1==5 || #1==6}\ifnum\pgfmathresult>0\relax%
	\fill[black] (0.5,0.5) circle[radius=1/6];    
	\fill[black] (-0.5,-0.5) circle[radius=1/6];\fi 
	\pgfmathparse{#1==3 || #1==4 || #1==5 || #1==6}\ifnum\pgfmathresult>0\relax%
	\fill[black] (-0.5,0.5) circle[radius=1/6];    
	\fill[black] (0.5,-0.5) circle[radius=1/6];\fi 
	\pgfmathparse{#1==1 || #1==3 || #1==5}\ifnum\pgfmathresult>0\relax%
	\fill[black] (0,0) circle[radius=1/6]; \fi 
	\ifnum#1=6\relax%
	\fill[black] (0.5,0) circle[radius=1/6];     
	\fill[black] (-0.5,0) circle[radius=1/6];\fi 
}

\newcommand{\dice}[5]{
	\begin{scope}[shift={(#1mm,-#2mm)}, rounded corners=#5, fill=White, scale = 0.12]
		\begin{scope}[canvas is xy plane at z=-1]
			\filldraw (-1,-1) rectangle (1,1);
		\end{scope}
		\begin{scope}[canvas is xz plane at y=-1]
			\filldraw (-1,-1) rectangle (1,1);
		\end{scope}
		\begin{scope}[canvas is yz plane at x=-1]
			\filldraw (-1,-1) rectangle (1,1);
		\end{scope}
		\begin{scope}[canvas is xy plane at z=1]
			\filldraw (-1,-1) rectangle (1,1);
			\dicenum{5}
		\end{scope}
		\begin{scope}[canvas is xz plane at y=1]
			\filldraw (-1,-1) rectangle (1,1);
			\dicenum{1}
		\end{scope}
		\begin{scope}[canvas is yz plane at x=1]
			\filldraw (-1,-1) rectangle (1,1);
			\dicenum{3}
		\end{scope}
	\end{scope}
}

\begin{figure}[h]
	\centering	
	\begin{tikzpicture}	
		\slot{1}{0}{0}
		\slot{2}{12}{0}
		\slot{3}{24}{0}
		\node[dot, shift={(34mm,-0mm)}](dot1){\textbf{\dots}};
		\node[slot,shift={(44mm,-0mm)}, fill = Red!10] (slot32) {slot \\[-3pt] 32};
		\node[left = 0mm of slot1,rotate=45,anchor = south east,yshift=2mm,xshift=2mm](prop){Proposer};
		\node[check, shift={(0mm,-10mm)}] (c1) {\textbf{\cmark}};
		\node[check, shift={(24mm,-10mm)}] (c3) {\textbf{\cmark}};
		\node[xmark, shift={(12mm,-10mm)}] (x2) {\textbf{\xmark}};
		\node[qmark, shift={(44mm,-10mm)}] (q32) {\textbf{?}};
		\node[dot, shift={(34mm,-10mm)}](dot2){\textbf{\dots}};
		\dice{0}{19}{0}{0}{0}
		\dice{24}{19}{0}{0}{0}
		\node[base,shift={(12mm,-19mm)}] (zero1) {0x00};
		\node[dot, shift={(34mm,-19mm)}](dot3){\textbf{\dots}};
		\node[base,shift={(6mm,-19mm)}] (xor1) {$\oplus$};
		\node[base,shift={(18mm,-19mm)}] (xor2) {$\oplus$};
		\node[base,shift={(30mm,-19mm)}] (xor3) {$\oplus$};
		\node[base,shift={(38mm,-19mm)},inner sep=-1pt] (xor4) {$\oplus$};
		\dice{44}{23}{0}{0}{0}
		\node[base,shift={(44mm,-19mm)}] (or) {or};
		\node[base,shift={(44mm,-16mm)}] (zero2) {0x00};
		\node[left = 0mm of c1,rotate=45,anchor = south east,yshift=2mm,xshift=2mm](rev){Revealed ?};
		\node[check, fill = White, shift={(0mm,-19mm)},opacity=0] (ghost) {\textbf{\cmark}};
		\node[align = center, left = 0mm of ghost,rotate=45,anchor = south east,yshift=1mm,xshift=1mm](RANDAO){RANDAO \\[-4pt] mix};
		
		\draw[Blue] (slot1) -- (c1);
		\draw[Blue] (slot2) -- (x2);
		\draw[Blue] (slot3) -- (c3);
		\draw[Blue] (slot32) -- (q32);
		\node[shift={(0,-18.3mm)}] (ghost2) {};
		\draw[Blue,-{Stealth[length=1.5mm]}] (c1) -- (ghost2);
		\draw[Blue,-{Stealth[length=1.5mm]}] (x2) -- (x2|-ghost2.north);
		\draw[Blue,-{Stealth[length=1.5mm]}] (c3) -- (c3|-ghost2.north);
		\draw[Blue,-{Stealth[length=1.5mm]}] (q32) -- ($(zero2.north)+(0mm,-2.5mm)$);
		
		\node[above right = 1mm of slot32, text = Mahogany,rectangle, inner sep =1pt, rotate=-20] (attack) {Attacker};
		\draw[Mahogany] (attack)--(slot32);
		\node[right = 2 mm of zero2, align = center, text = Mahogany, inner sep = 1pt] (seed1) {seed \\[-3pt] n°1};
		\node[below = 1 mm of seed1, align = center, text = Mahogany, inner sep = 1pt] (seed2) {seed \\[-3pt] n°2};
		\draw[Mahogany,-{Stealth[length=1.5mm]}] ($(seed1.west)+(-3mm,0)$)--(seed1);
		\draw[Mahogany,-{Stealth[length=1.5mm]}] ($(seed2.west)+(-3mm,0)$)--(seed2);
		\draw[Mahogany,-{Stealth[length=1.5mm]}] ($(xor4.north)$)--($(zero2.west)+(-0mm,+0mm)$);
		\draw[Mahogany,-{Stealth[length=1.5mm]}] ($(xor4.south)$)--($(zero2.west)+(1.5mm,-7mm)$);
	\end{tikzpicture}
	\caption{Illustration of Ethereum RANDAO LRA}
	\label{fig:LRA}
	\Description{A depiction of a Last Revealer Attack on Ethereum's RANDAO. Each proposer includes a random input in their proposed blocks (that is the revealed value of a previous commit). When a block is not proposed, the input defaults to 0. The result of RANDAO is the XOR of those inputs, this opens up the possibility to manipulate the output if someone is supposed to be the last revealer: Two resulting random seeds can be computed before proposing, one resulting from normally proposing a block, one resulting from withholding the block}
\end{figure}
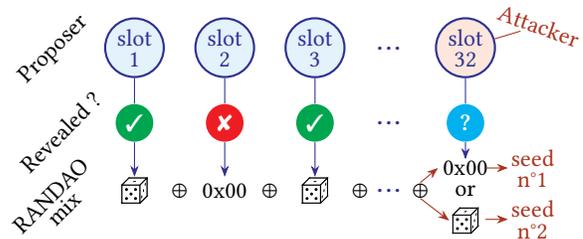

To clarify the LRA, we examine the revelation phase (see Figure~\ref{fig:LRA}). During an epoch, block proposers must include their revealed secret numbers, or else their blocks are ignored. Thus, each slot results either in a block with the revealed value (green check-mark), or no block (red "x"). The resulting random seed, computed by XOR-ing revealed values (missing reveals default to 0), determines block proposal rights for the \emph{next} epoch (of the same parity). The last revealer can decide whether to propose (revealing their secret, yielding seed n°2) or withhold (yielding seed n°1). The attacker then selects the seed yielding the most favorable block allocations. As detailed in~\cite{Randaomanip}, several incentives motivate this choice; here, we focus solely on rewards from accepted block proposals, normalized to $1$ per block. One must also take into consideration that withholding a block results in an opportunity cost of $1$ since no reward will be gained. Consider this simple example:
\begin{example}
	The agent can either 1) propose a block, yielding random seed n°1 with $2$ block proposal rights in the next epoch, or 2) withhold the block (possibly disguised as a network issue), yielding random seed n°2 with $5$ proposal rights next epoch. Although strategy 2) costs $1$ block proposal immediately, it ultimately yields $3$ additional rights in the following epoch, making it the rational choice.
\end{example}

Now suppose the attacker controls the final $2$ blocks of an epoch. He can choose between proposing both, hiding then proposing, proposing then hiding, or hiding both—giving him a choice among $4$ random seeds. More generally, controlling the last $k$ blocks provides a choice among $2^k$ seeds. While prior analyses typically stop at this first-order attack~\cite{Randaomanip}, controlling the final blocks in epoch $e$ enhances opportunities in epoch $e+1$, making these blocks inherently more valuable. Thus, the attacker faces a decision problem: maximize immediate rewards or strategically position himself to increase future block control and rewards. Consider this example:
\begin{example}
	The agent controls the last block of the epoch. If he 1) proposes, he receives 2 proposal rights next epoch (including final slots); if he 2) hides, he receives 4 rights but none at epoch's end. These choices yield net rewards of $2$ and $3$, respectively. However, strategy 1) offers control over the next epoch’s final blocks, possibly justifying a short-term loss if expected future rewards exceed $1$.
\end{example} 

\subsection{Example: A simple card game}\label{sec:cardGame}

We introduce a fictional card game designed to illustrate the above mechanism, it makes the concepts discussed throughout this paper more intuitive. Table~\ref{tab:cardgame} illustrates the game. We assume an infinite number of standard decks shuffled together, ensuring each card draw is uniform and IID (identical cards can appear consecutively). Each round proceeds as follows:
\begin{itemize}
	\item The dealer draws $d$ cards (the first round, $d=2$)
	\item The player chooses card $i$
	\item He earns the value of the card (figures are valued at 10)
	\item He pays $7$ if he picked the first card, $8$ if he picks the second one, and so on
	\item The dealer clears the cards, and depending on the color of the picked card (club, spade, diamond, heart) we respectively set $d=1,2,4,8$ for the next round.  
\end{itemize}

\begin{table}[t]
	\centering
	\caption{Card game: a round with 4 cards}
	\label{tab:cardgame}
	\begin{tabular}{rcccc}
		\toprule
		Card \#$i$ & 1 & 2 & 3 & 4\\
		\midrule
		Cost $c(i)$ & 7 & 8 & 9 & 10\\
		Cards
		& \includegraphics[width=10mm,keepaspectratio]{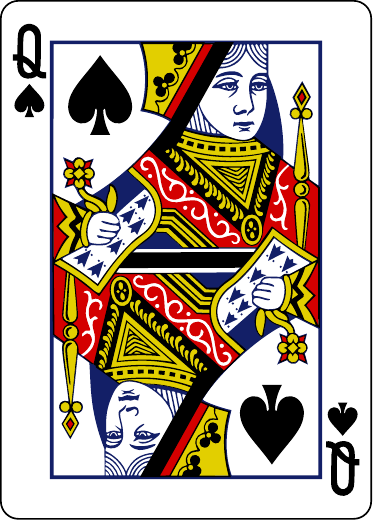} & \includegraphics[width=10mm,keepaspectratio]{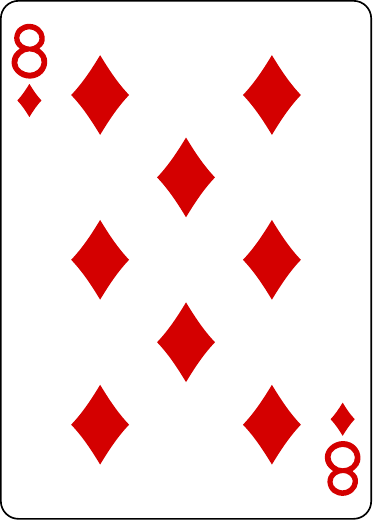}&  \includegraphics[width=10mm,keepaspectratio]{8K.pdf} &  \includegraphics[width=10mm,keepaspectratio]{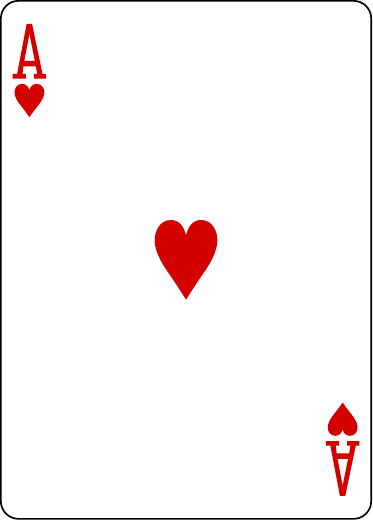}\\
		Reward $r$ & 10 & 8 & 8 & 1 \\
		Net reward $R$ & 3 & 0 & -1	 & -9\\
		\# draws next round & 2 & 4 & 4 & 8\\
		\bottomrule
	\end{tabular}
\end{table}

Representing this problem as an MDP requires capturing the full state information before actions are chosen. With up to 8 cards on the table, there are more than $40^8$ possible states ($40$ instead of $52$ since all face cards count as 10). Given $d\in{1,2,4,8}$, this results in $40 + 40^2 + 40^4 + 40^8$ total states, causing a combinatorial explosion typical of MDP formulations.

\section{Modeling}
We now present our notations for MDPs, and then introduce \emph{pseudo MDPs} (pMDPs) and explain the two problem reductions (ex-post and ex-ante). We use the notations $\setN$, $ \setNs $, $\setNn{n}$ and $\setNsn{n}$ for the sets of natural integers with/without $0$ (with the $^*$), and bounded by $n$ (with $_ n$).

\subsection{Markov Decision Process}
\begin{definition}\label{def:MDP}
	An MDP is a tuple $(\setS,\setA,P,R)$, where when the agent is in state $s\in \setS$, he picks an action $a\in \setAs$. The next state is $s'\in \setS$ with transition probability $P(s'|s,a)$ while a reward $R(s,a)$ is granted to the agent.
\end{definition}

\paragraph{Card game's case:} \label{parag:cardgameMDP} 
To describe the game with an MDP we represent a card by its color-value pair $(\sigma,r)$. 
Noting $\setS_o$ the set of possible $(\sigma,r)$, we have $|\setS_o|=40$ since the figures are all valued $10$. We then define $\setS = \bigcup_{d\in\{1,2,4,8\}} \setS_o^d$. 
For a state $s\in \setS$ we note $d(s)$ the number of cards of $s$, the notation $d$ is overloaded as $d(\sigma)$ also represent the number of cards drawn next turn depending on the picked color $\sigma$. 
Additionally the state $s$ can be decomposed into its color-value pairs $s=(s_i)_{i\in\setNsn{d(s)}}$ with for each $i$: $s_i=(s_i^\sigma,s_i^r)$. 
We can then define $\setAs = \setNsn{d(s)}$. 
For each $(s,a,s')\in \setS\times\setAs\times\setS$ 
we have:
\[ P(s'|s,a) = \ind{d(s')=d(s_a^\sigma)} \prod_{i\in\setNsn{d(s')}} P(s'_i)  \]
with $\ind{.}$ 
being the indicator function and for each $s'_i\in\setS_o$ 
the probability $P(s'_i)$ is $1/52$ if $s'^r_i<10$ and $4/52$ otherwise 
(for each color, 4 cards are valued 10). This comes from the fact that when picking action $a$ from state $s$ we know that we will draw $d(s_a^\sigma)$ cards, then those $d(s')$ cards are drawn IID. 
Finally, for $(s,a)\in\setS\times\setAs$, we would have a net reward $R(s,a) = s^r_a-c(a)$ (here $c(a)$ represents the cost of choosing card $a$, we have $c(a)=6+a$); additionally, we will note this net reward $s^R_a = R(s,a)$.
This would be the natural way to represent the card game as an MDP, it is rather complex and the size of $S$ could prevent its solve though standard dynamic programming solutions.
Note that in the rest of the paper we keep using $r$ as a reward from a card value and $R$ as a net reward taking into account the cost $c$.

\subsection{Pseudo Markov Decision Process}\label{sec:pMDP}

Before the drawing of the cards for a new round, when we already know $d$, we could consider ourselves as being in an \emph{ex-ante state}. pMDPs provide a framework describing our problems more intuitively from this ex-ante state. 
\begin{definition}\label{def:pMDP}	
	A pMDP is a tuple $(\setSig,\setR,P,d,c)$: Starting from the ex-ante state $\sigma\in\setSig$, we realize an IID draw of $d: \setSig \rightarrow \setNs$ pairs $(\sigma'_i,r_i)_{i\in\setNsn{d(\sigma)}} \in \setS_\sigma$  with probabilities $P(\sigma',r|\sigma)$ where $\setS_\sigma =\setS_o^{d(\sigma)}$ with $ \setS_o = \setSig \times \setR$. The agent then picks an outcome $i\in\setNsn{d(\sigma)}$, pays the cost $c(i)$ and earns $r_i$; The next ex-ante state is $\sigma'_i$. 
\end{definition}

We overload $ P $ to represent different probability functions and distributions, context and variables prevent ambiguity. 

\paragraph{Card game's case:} $\sigma$ is the color of the previously picked card ($\setSig=\{$club, spade, diamond, heart$\}$); 
For rewards $\setR = \setNsn{10}$; 
As before, the uniform draw of a card (over the 52 cards) yields $P(\sigma',r|\sigma)$, that is $4/52$ if $r=10$ and $1/52$ otherwise ; 
the picked card's color leads to a state $\sigma'\in\setSig$. 
The function $d$ respectively associates $1,2,4,8$ to club, spade, diamond, heart and the cost is $c(i)= 6 + i$. Already we can see how much more naturally pMDPs model such a problem.

\begin{table*}[t]
	\caption{Problem sizes comparison ex-ante vs ex-post MDP reductions}
	\label{tab-pmdpReducSizeCompare}
	\centering
	\begin{tabular}{lrrrr}
		\toprule
		Set Cardinal & $ |\setS| $ & $ \max_s|\setAs| $ & $ |\setSig| $ & $ \max_\sigma|\setAlph(\sigma)| $ \\
		\midrule
		Card Game (numerical) & $ \approx 40^8 $ & $ 8 $ & $ 4 $ & $ 8^{40^8} $ \\
		LRA on RANDAO & $ \bigO{\kappa^{2^{\kappa+1}}} $ & $ 2^\kappa $ & $ \kappa+1 $ & $ \bigO{2^{{\kappa^{2^{\kappa+1}+1}}}} $\\
		General Case  & $ \sum\limits_{d^\prime \in d(\setSig)} (|\setSig|\,|\setR|)^{d^\prime} $ & $ \max d $ & $ |\setSig| $ & $ (\max d)^{(|\setSig|\,|\setR|)^{\max d}} $ \\
		\bottomrule
	\end{tabular}
\end{table*}

\subsection{Reducing pMDPs to MDPs}\label{sec:reduce}
We propose two reductions from a pMDP to an MDP. The first, more natural approach—used in the card game example—returns to the standard MDP setup where states reflect the situation \emph{after} randomness: the \emph{ex-post MDP}. The second keeps the smaller \emph{ex-ante} state set as the MDP’s state space. \textbf{Both} reductions are essential to the understanding of the remainder of the paper.

\paragraph{Ex-post MDP:} For a pMDP $(\setSig,\setR,P,d,c)$, we set the state set as the possible ex-post joint draws: $\setS=\left(\setSig\times\setR\right)^{d(\setSig)}$, for $s\in  \setS$, the action set is $\setAs=\setNsn{d(s)}$. 
For  $(s,a,s')\in \setS\times\setAs\times\setS$, since for pMDPs the random process generating $s'$ only depends on $\sigma'=s^\sigma_a$ 
(we use the same notations as in the card game MDP in section~\ref{parag:cardgameMDP})
, the joint probability of drawing the $(s'^\sigma_i,s'^r_i)$ pairs from $s^\sigma_a$ is $P(s'|s,a) = P(s'|s^\sigma_a) = \prod_{i\in \setNsn{d(s')}} P(s'^\sigma_i,s'^r_i|s^\sigma_a)$ if $ d(s') = d(s_a^\sigma) $, and $ 0  $ otherwise. 
Finally we set $R(s,a) = s^R_a = s^r_a-c(a)$. 
The MDP $(\setS,\setA,P,R)$ is the ex-post reduction.

\paragraph{Ex-ante MDP:} For a pMDP $(\setSig,\setR,P,d,c)$, we keep the state space $\setSig$. 
The action set is the partial strategies for ex-post states accessible from $\sigma\in\setSig$, that is: $ \setAsig = \{ \alpha: s\in \setS_\sigma \mapsto s_i\in\setS_o,\, i\in\setNsn{d(\sigma)}\}$. 
For $\alpha\in\setAsig$, we note $\alpha^\sigma$ and $ \alpha^r$ the projections of $\alpha$ on $\setSig$ and $\setR$, and $\alpha^R$ the cost adjusted version of $\alpha^r$. 
For $ (\sigma,\alpha,\sigma') \in \setSig\times\setAsig\times\setSig $ the transition probability is $P(\sigma'|\sigma,\alpha) = \sum_{s\in \setS_\sigma} P(s|\sigma) \ind{\alpha^\sigma(s)=\sigma'}$ with the joint probability $P(s|\sigma) = \ind{d(s) = d(\sigma)}\prod_{i\in\setNsn{d(s)}} P(s^\sigma_i,s^r_i|\sigma)$ (details in footnote\footnote{
We decompose over the possible $s$ reached between $\sigma$ and $\sigma'$, this gives us:  $P(\sigma'|\sigma,\alpha) = \sum_{s\in \setS_\sigma} P(\sigma',s|\sigma,\alpha) = \sum_{s\in \setS_\sigma} P(s|\sigma,\alpha) P(\sigma'|s,\sigma,\alpha)$. Then we use the fact that the drawing of $s$ only depends on $\sigma$ and not $\alpha$ to have  $ P(s|\sigma,\alpha) = P(s|\sigma) $; We also use the fact that from $s$, the transition to $\sigma'$ is deterministic and needs $\sigma' = s_a^\sigma$. And given $s$ and $\alpha$ we have $ s_a^\sigma = \alpha^\sigma(s) $. Those two points yield: $P(\sigma'|\sigma,\alpha) = \sum_{s\in \setS_\sigma} P(s|\sigma) \ind{\alpha^\sigma(s)=\sigma'}$.
}). 
Finally we set $\mathrm{R}(\sigma,\alpha)=\esp(\alpha^R(s)|\sigma)$, for the same joint probabilities $P(s|\sigma)$. 
This gives us the ex-ante MDP: $(\setSig,\setAlph,P,\mathrm{R})$.

\paragraph{Set sizes:} 

Table~\ref{tab-pmdpReducSizeCompare} provides a comparison of the different set sizes. We can see that in the ex-post reduction, the complexity is absorbed by the state set $\setS$, while in the ex-ante reduction, it is in the action set $\setAlph$. Note that for Ethereum's RANDAO, since $\kappa=32$, the ex-post state size is of the order of $10^{10^{10}}$,  this makes the problem completely intractable for standard dynamic programming algorithms. For comparison, the go game state space is bounded by $3^{19^2}$. Additionally, the actions available, in the worst case of the attacker controlling the 32 blocks of the epoch would be described by $2^{32}$ pairs of $(\sigma^\prime, r)$, that is a roughly $5 \mathrm{GB}$ input size.

\section{Value Iteration and Policy Extraction}

Here, we present value iteration a fundamental algorithm used to solve MDPs. 
It is based on iterating the Bellman operator (equation~\ref{equ: Bellman Operator}, see~\cite{Bellman1957} for details) to find state values. 
With the state values, \emph{Policy Extraction} provides an optimal policy.
We look at how those Algorithms can be adapted to reduced pMDPs, increasing their performance. 
Furthermore, we provide necessary conditions extending those results to a larger sub-class of MDPs. 
Finally, we show how the approach can be used in Monte Carlo Value Iteration (MCVI). 

\subsection{MDPs}\label{sec:solveMDP}
\paragraph{Value Iteration:} (pseudocode in appendix~\ref{app: algorithms}, algorithm~\ref{alg:ValueIteration}), computes a \emph{value function} which assigns a value to each state when following a given policy $\pi:s\in\setS \mapsto a\in\setAs$. 
This value is the expected actualized sum (for a given  a \emph{discount factor} $  \gamma $) of the rewards an agent gets when starting in that state: $ V_\pi(s) = \esp\left[\sum_{t=0}^\infty \gamma^t R_t|s_0=s, \pi \right] $ where $s_t$ and $R_t$ are the state and reward at time $t\in\setN$. An optimal policy $\pi^*$, produces the optimal value function $ V^*(s) $. 
The Bellman Operator $ T^*$ admits $ V^* $ as its fixed point: 
\begin{align}\label{equ: Bellman Operator}
	T^*V(s) =  \max_{a\in\setAs} \left[ R(s,a)+ \gamma \esp(V(s')|s,a) \right] 
\end{align} 
This operator being contracting, iterating it over any initial $V$ converges exponentially fast to $V^*$. However, the complexity (see appendix~\ref{app: algorithms}) of an iteration is in $\bigO{|\setS|^2 |\setA|}$ which can make the algorithm intractable for large problems. Recall that for pMDPs $|\setS|\in\bigO{(|\setSig|\, |\setR|)^{\max d}}$ since in RANDAO's LRA case it means $|\setS|\in\bigO{\kappa^{2^{\kappa+1}}}$, any algorithm complexity containing $|\setS|$ is not usable in our case. In the rest of the paper we also use the fact that $|\setA|= {\max d}$ for pMDPs.

\paragraph{Policy extraction} (pseudocode in appendix~\ref{app: algorithms}, algorithm~\ref{alg:Policy Derivation}).
In general, once the optimal value function $V^*$ of an MDP $(\setS,\setA,P,R)$ is known, an optimal policy $\pi^*$ can be computed with a complexity for a single state $s\in\setS$ in $\bigO{|\setS|\, |\setA|}$ (see appendix~\ref{app: algorithms}):
\begin{align}\label{equ: Policy Extraction}
	\pi^*(s) = a^*\in\argmax\limits_{a\in\setAs} \left[ R(s,a)+ \gamma \esp(V^*(s')|s,a) \right] 
\end{align}

\subsection{Reduced pMDPs}
We use the previous notations for both reductions ex-post $(\setS,\setA,P,R)$ and ex-ante $(\setSig,\setAlph,P,\mathrm{R})$. 
In the ex-ante reduction, the sets $\setAlph$ being sets of functions, their sizes are enormous and a naive value iteration would yield a terrible result. 
However, instead of iterating over the possible functions $\alpha$ (Bellman operator~\ref{equ: Bellman Operator}'s $ \max $), we iterate over the possible outputs ($a\in\setAs$), input by input ($s\in\setS$) to reach the $\max_\alpha$. The intuition behind this can be fragmented in seeing optimal ex-ante and ex-post value functions as functions of one another. Intuitively, we have the two half Bellman equations:
\begin{equation}\label{eq:ValueFunc1}
	W^*(\sigma) = \esp_{s\sim P(s | \sigma)} \left[V^*(s)\right]
\end{equation}
\begin{equation}\label{eq:ValueFunc2}
	V^*(s) = \max_{a\in\setAs}\left[s^R_a + \gamma W^*(s_a^\sigma)\right]
\end{equation}
Combining those equations results in a better way to write the Bellman operator for reduced pMDPs, yielding a better value iteration (pseudocode in appendix~\ref{app: algorithms}, algorithm~\ref{alg:ValueIterationExAnteReducedpMDP}). 
Noting $W(\sigma)$ the value function for ex-ante state in the ex-ante MDP, the Bellman operator becomes (proof in appendix~\ref{app:BellmanOperatorExAntepMDP}):
\begin{align}
	T^*W(\sigma) & = \esp\limits_{s\sim P(s|\sigma)} \left[ \max_{a\in\setNsn{d(\sigma)}} R(s,a ) + \gamma W(s_a^\sigma) \right]\label{eq:BellmanExAnteMDP}
\end{align}
The computation's complexity is in $ \bigO{|\setSig|\, |\setS|\, \max d} $ (see algorithm~\ref{alg:Policy Derivation exAnte reduced pMDP} in appendix~\ref{app: algorithms}) which reduces the iteration complexity by a factor of $ \lambda = |\setS|/|\setSig|  $ from the natural MDP (the ex-post MDP). In the card game case $ \lambda > 40^8/4 $, while for the LRA $\lambda$ is of the order of $ \kappa^{2^{\kappa+1}-1}$, this is gigantic, yet still entirely insufficient. 

\paragraph{Policy Extraction} This result uses a point of view in between the ex-ante and ex-post reductions. Its advantage comes from the fact that reduced pMDPs have either a large state space $\setS$ (ex-post reduction) or a large action space $\setAlph$ (ex-ante reduction). The hybrid view allows us to better circumvent the scanning of those sets. It can be extended to Policy Extraction where given the ex-ante state values $W$, one can compute an optimal action in an ex-post state $s \in \setS$ : 
\begin{equation}\label{equ:fastOptimalPolicyPMDP}
	\pi^*(s) = a^*\in\argmax\limits_{a\in\setNsn{d(\sigma)}} \left[ s^R_a + \gamma W^*(s^\sigma_a) \right]
\end{equation}
Once in $s$, the agent only considers actions in $\setAs$ , and only needs to know $W$ which is of size $\setSig$. Extracting an optimal action has a time complexity in $\bigO{d(s)}$ and a space one in $ \bigO{|\setSig|} $. In the card game example: $|\setSig|=4$ and $|\setAs|\le8$, this is extremely economic in time and space, even a human could implement this\footnote{In fact, casino card counters use a similar trick, they only maintain a condensed information about the state of the dealer's deck (the ex-ante state), and adapt their action $a\in\setAs$ accordingly within a large ex-post state set using simple decision rules}. This is also extremely useful for embedded systems with strong computation and memory constraints.

Equations~\ref{eq:BellmanExAnteMDP}, \ref{equ:fastOptimalPolicyPMDP} give this theorem (proof is in appendix~\ref{app:BellmanOperatorExAntepMDP}):

\begin{theorem}\label{thm: exAnte MDP value iteration}
	Given a pMDP  $(\setSig,\setR,P,d,c)$, and its reduced MDPs $(\setS,\setA,P,R)$ and  $(\setSig,\setAlph,P,\mathrm{R})$. There exist a dynamic programming implementation of value iteration of the ex-ante reduction whose complexity is in $ \bigO{|\setSig|\, |\setS|\, \max d} $ using the Bellman operator as in equation~\ref{eq:BellmanExAnteMDP}.
	One can then extract an optimal action in $\bigO{\max d}$.
\end{theorem}
Note that given the size of $\setS$, see Table~\ref{tab-pmdpReducSizeCompare}, this is not satisfactory yet for RANDAO's LRA.

\subsection{Necessary conditions for standard MDPs}\label{sec:necessary conditions}

We propose to enlarge the previous results to a larger sub-class of MDPs:
\begin{theorem}\label{th: XA MDP}
	Given an MDP $(\setS,\setA,P,R)$, if there exists a random variable $ \sigma' $ with support $\Sigma $ and whose realization is dependent solely on $s$ and $ a $ with distributions $P(\sigma'|s,a)$ for all $ \sigma'\in\Sigma, s\in\setS, a\in\setAs $ such that $s'$ is conditionally independent given $\sigma'$. 
	
	Then we can define a second MDP whose state set is $\Sigma$ and whose optimal value function is $W^*$, additionally we have:
	\begin{equation}\label{equ: Bellman XA MDP}
		T^*W(\sigma) = \esp_{s\sim P(s|\sigma)} \left[ \max_{a\in\setAs} R(s,a ) + \gamma\, \esp(W(\sigma')|s,a) \right]
	\end{equation}
	and for $s\in\setS$:
	\begin{equation}\label{equ: Policy extraction XA MDP}
		\pi^*(s) = a^*\in\argmax\limits_{a\in\setAs} \left[ R(s,a)+ \gamma \esp(W^*(\sigma')|s,a) \right]
	\end{equation}
	Computations are respectively in $ \bigO{|\setS|\, |\setSig|^2\,  |\setA|} $ and $ \bigO{|\setSig|\, |\setA|} $
\end{theorem}
The proof is in appendix~\ref{app:proofTHMPolicyDerivation}, and is similar to that of Theorem~\ref{thm: exAnte MDP value iteration}. The difference in complexity comes from the fact that where we could directly access $W(\sigma')$ in reduced pMDP thanks to the deterministic transition from $(s,a)$ to $\sigma' = s_a^\sigma$, we now have to compute $\esp(W(\sigma')|s,a)$. The adapted pseudocode is in appendix~\ref{app: algorithms}, algorithms \ref{alg:ValueIterationMDPwithCond} and \ref{alg:Policy Derivation exAnte}.
The needed hypothesis represents an intermediate step during the Markov transition: when in state $s\in\setS$ with $a\in\setAs$ chosen, we first draw $\sigma'\in\setSig$ with $P(\sigma'|s,a)$. 
This $\sigma$ contains all the information needed to draw $s'\in\setS$ (with probability $P(s'|\sigma')$). This structure is exactly that of repeated two-stage stochastic control problems where the $\sigma$-states represent possible first-stage situations and the $s$-states represents the second stage (often named \textit{recourse}, see~\cite{twoStage}).

\subsection{Monte Carlo Value Iteration (MCVI)}

Reduced pMDPs and MDPs from theorem~\ref{th: XA MDP} can use the values $ W $ of the ex-ante states as a heuristic which can be used, for instance, in an adapted Monte Carlo Value Iteration (MCVI)\footnote{
	MCVI comes in different flavors: for each sampled states, actions could be sampled with an exploratory policy. In our case we are exhaustive over actions. We focus on estimating expected values with sampling}.
Here, we sample the ex-post states $s$ from each possible ex-ante state $\sigma$. In each of the sampled ex-post state $s$, the value $V(s)=\max_{a\in\setAs} R(s,a) + \gamma \esp(W(\sigma')|s,a) $ is computed and then averaged to update $W(\sigma)$. 
The pseudocode can be found in appendix~\ref{app: algorithms}, algorithm~\ref{alg:MCVI MDP from thm} , it leads to an iteration complexity of $\bigO{|\setSig|^2\, |\setA|\, n_{sample}\, \varsigma}$ where $\varsigma$ represents the sampling cost and allows one to bypass scanning $ \setS $ altogether. The cost of sampling for MDPs is usually in $ \bigO{1} $ but for pMDPs sampling $s$ naturally from $\sigma$ requires $d(\sigma) = d(s) = |\setAs|$ individual sampling of $(\sigma',r)$. While this works perfectly for the card game, where $\max d = |\setA| = 8$, for the LRA we have an issue since $\max d = 2^\kappa$, all in all this gives for the LRA a complexity in $ \bigO{\kappa^2\,2^{2\kappa}\, n_{sample}}$. This is still an issue, especially if we want to investigate larger values for $\kappa$, additionally, with MCVI, we lose the convergence guarantees.

\section{Value Iteration with Variable Change}\label{sec:VariableChange}

The motivation for the following work comes from the fact in ex-ante reduced pMDPs, the state values $W^*(\sigma)$ have a particular relationship to $d(\sigma)$. 
In the card game, for example, the only reason $W^*(club)$ would be different from $W^*(spade)$ is the fact that $d(club)=1$ while $ d(spade)=2 $. It follows naturally that $W^*(club) < W^*(spade)$ as it offers less choice, and more than that, rewriting equation~\ref{eq:ValueFunc1} gives: $ W^*(club) =  \esp_{s\sim P(s | club)}\left[V^*(s)\right]$ and intuitively: $ W^*(spade) =  \esp_{s_1,s_2\sim P(s | club)}\left[\max (V^*(s_1), V^*(s_2))\right]$. Note that here the sampling is made in the same way, it is just doubled while the better outcome is picked in the $\esp$.
This section explores how we can leverage this relationship between $W^*$ and $d$ to improve the computation of the Bellman operator $T^*$.

\subsection{Changing the random variable:} 
In general, the Bellman equation expresses the static relation between the value of a state (the state value function $V$) and the expected value for the agent after a single stochastic transition (composed of the reward and the value function of the next state $r + \gamma V$). We already used the dual reductions to improve our computation of the Bellman operator (see equation~\ref{eq:BellmanExAnteMDP}), relying on the formulation of equation~\ref{eq:ValueFunc1}; we now want to leverage the IID draws to better compute the ex-post value function from as seen in equation~\ref{eq:ValueFunc2}. 

Value functions represent a utility, and in equation~\ref{eq:ValueFunc2}, in ex-post state $s\in\setS$, the agent following an optimal policy will pick the best draw $s_a$ with $a\in\setAs$. This means that the agent will maximize the utility $u_W(s,a):=s_a^r -c(a) + \gamma\, W(s^\sigma_a)$ associated with a certain unitary draw given his perception of the ex-ante state value function $W$. When $W^*$, this leads to an optimal action choice. We note $u_W^*(s) := \max_{a\in\setAs} u_W(s,a)$ the ex-post maximal utility, this can be seen as a way to compute the ex-post state value function $V$ of the ex-post state $s$ for an a priory ex-ante state value function $W$.  Combined with equation~\ref{eq:ValueFunc1} this allows us to rewrite the ex-ante Bellman operator from equation~\ref{eq:BellmanExAnteMDP} as:

\begin{equation}\label{eq:BellmanVariableChange}
	T^*W(\sigma) = \esp_{s\sim P(s|\sigma)} \left[u_W^*(s) \right]\text{, with $ u_W^*(s) = \max_{a\in\setNsn{d(s)}} u_W(s,a) $}
\end{equation}

Instead of seeing $s\sim P(s|\sigma)$ as the random variable from the ex-ante – ex-post transition, we will consider the realization of $u_W^*(s)$, written as $u$ as being the new random variable. This means that the computation of the Bellman operator can be done by computing the expected utility $u$:

\begin{equation}\label{eq:BellmanVariableChangeUtility}
	T^*W(\sigma) = \esp_{u\sim P^W(u|\sigma)} \left[u\right]
\end{equation}
 
The ex-post optimal utility $ u $ is \emph{realized} by the stochastic transition between ex-ante state and ex-post state. It is a \textbf{max} of several $u_W(s,a)$ that are IID random variables once adjusted for the cost function's shift. This means that for a prior ex-ante state value function $W$ and for a given  ex-ante state $\sigma$, and actions $a_0,a_1 \in\setA$, we have a link between the distributions of the realized utilities $ u_W(s,a) $  of those actions: $P(u_W(s,a_0) + c(a_0)|\sigma) = P(u_W(s,a_1) + c(a_1)|\sigma)$. We note $\setU$ the set of value $u$ can take (its support). Given that each utility $u$ is realized for a certain ex-post state $s$ and action $a$, with the computation of $s_a^r-c(a)+\gamma W(s^\sigma_a)$, we know that $\setU \subset \setR - c(\setA) + \gamma W(\setSig)$ so $ |\setU| \le |\setSig|\,|\setR|\,|c(\setA)| = |\setSig|\,|\setR|\,\max d $
(recall that for reduced pMDPs, $\setA = \setNsn{\max d}$). 

For the card game, this bound is $4\times10\times8=320$ which is manageable, but $|\setU|$ is even smaller as some costs and rewards will yield the same utility: for instance $r=5$ and $c(i)=8$ will give the same utility as $r=4$ and $c(i) = 7$. 
Since $r-c(i) \in [-13,\dots,3]$ (size 17), we know that $ |\setUs|\le 68 $. 

In general, when the costs are similar to the rewards (in spread and support size): $|\setU|\in\bigO{|\setSig|\,|\setR|}$. For the LRA, since $|\setSig| =\kappa+1$ and since rewards and costs are both integers, and both are in $ \setNsn{\kappa} $, we have  $ |\setU| \in \bigO{\kappa^2}$. 

Now that the variable is changed, we only need its distribution $ P^W(u|\sigma) $ to be able to compute the Bellman operator as in equation~\ref{eq:BellmanVariableChangeUtility}. This is shown formally in appendix~\ref{app:variableChange}, and this is a simple computation in $ \bigO{|\setU|}$ which for our cases is in $\bigO{|\setSig|\,|\setR|}$ or in $\bigO{\kappa^2}$ as just discussed. 

For a prior ex-ante state value function $W$, the distribution of $u$ can be computed through brute force with its definition by enumerating all possible ex-post states $s\in\setS$, that is: $ P^W(u|\sigma) = \sum_{s\in\setS} P(s,u|\sigma) =  \sum_{s\in\setS} P(s|\sigma)P(u|\sigma,s)  $. Additionally we can write that: $P(u|\sigma,s) = \ind{u_W^*(s)=u}$ because $u$ is deterministic given $s$. Therefore, we can compute the distribution of the utility $u$ with: $ P^W(u|\sigma) = \sum_{s\in\setS} P(s|\sigma) \ind{u_W^*(s)=u} $. But this naive computation, which can be found in appendix~\ref{app: algorithms}, algorithm~\ref{alg: Naive PW distribution} would yield a complexity in $ \bigO{|\setSig|\,|\setS|\,\max d} $. With it, we do not reach any improvement. 

\begin{table*}[t]
	\caption{Relative ex-ante state values of the two problems: Card Game | RANDAO's LRA}
	\label{tab:resultValue}
	\begin{tabular}{lcccc|lcccc}
		\toprule
		$\Delta W_\sigma{(\sigma')}$ & $\sigma=c.$ &$\sigma=s.$ &$\sigma=d.$ &$\sigma=h.$ & & $\sigma=0$ &$\sigma=1$ &$\sigma=2$ &$\sigma=3$\\
		\midrule
		$\sigma'=club$ &0.00 & -1.41 & -2.01 & -2.07 & $\sigma'=0$ &0.00 & -1.16 & -2.12 & -2.95 \\
		$\sigma'=spade$ &1.41 & 0.00 & -0.60 & -0.66 & $\sigma'=1$ &1.16 & 0.00 & -0.97 & -1.79 \\
		$\sigma'=diam.$ &2.01 & 0.60 & 0.00 & -0.06 & $\sigma'=2$ &2.12 & 0.97 & 0.00 & -0.82 \\
		$\sigma'=heart$ &2.07 & 0.66 & 0.06 & 0.00 & $\sigma'=3$ &2.95 & 1.79 & 0.82 & 0.00  \\
		\bottomrule
	\end{tabular}
\end{table*}
\subsection{Efficient computation of the utility distribution:}
This is where we finally leverage the relationship between $ W^* $ and $d$ to improve the computation of $ P^W(u|\sigma)$.
We start by using cumulative distributions: $P^W(u\le\upsilon|\sigma)$ for $\upsilon\in\setU$. This allows us to use the properties of the distribution of the $\max$ function\footnote{Given two numerical random variables $X$ and $Y$,  and a number $z$, we have: $ P(\max(X,Y)\le z) = P(X\le z)P(Y \le z) $} in the Bellman operator. 
Finally, we introduce the notation $P^W_{i,j}(u\le\upsilon|\sigma) = P^W(\max_{i\le a <j} u_W(s,a) \le \upsilon|\sigma)$ representing the probability that $u$ is smaller than $\upsilon$ when restraining the agent's choice to draws of index $a$ with $1\le i\le a <j \le d(\sigma)+1$. With this we can realize a dichotomy on the computation of $P^W(u\le\upsilon|\sigma)$ since, in general:

\begin{equation}\label{equ:dichotomy}
	P^W_{1,i}(u \le \upsilon |\sigma) = P^W_{1,j}(u \le \upsilon |\sigma) P^W_{j,i}(u \le \upsilon |\sigma)
\end{equation}

This dichotomy however does not provide automatically a faster computation. Let us look at the first case where $j = i+1$. The only thing that makes $P^W_{1,2}(u \le \upsilon |\sigma)$ different from $P^W_{i,i+1}(u \le \upsilon |\sigma)$ is the cost $c(i)$ versus the cost of $c(1)$ (as mentioned in the previous section). This means that we can deduce one from the other only with an indexing change in $ \setU $:
\[ 
P^W_{i,i+1}(u \le \upsilon |\sigma) = P^W_{1,2}(u \le \upsilon + c(i) - c(1) |\sigma)
\]

Unfortunately, without additional assumptions, this only holds for extracted draws of size 1, that is when $j=i+1$. 
However, with certain cost functions, this works for any extraction size. For a constant cost, this is trivial since the distribution is the same. Let us consider the next function family that comes to mind: what if the cost is linear just like in the card game $c(i) = c_0 + c_1 i$. In this case, the change of the index of the draw extraction will just linearly change indexation in the probability:
\[ 
P^W_{i,j}(u \le \upsilon |\sigma) = P^W_{1,j-i+1}(u \le \upsilon + c_1 (i-1) |\sigma)
\] 

In general, we need a function $\varsigma:i\in \setN \mapsto \varsigma(i) \in \mathbb{R}$ such that the distributions over the sub-intervals can be deduced from each other. For instance when splitting $ \setNsn{2i} $ in $ \setNsn{i} $ and $ \setNsn{i} + i $, we need the function to allow us to shift the extraction interval $[i+1,\ 2i]$ in the following way:
\begin{equation}\label{equ: subintervals condition}
	P^W_{i+1,2i+1}(u \le \upsilon |\sigma) = P^W_{1,i+1}(u \le \upsilon + \varsigma(i) |\sigma)
\end{equation}
Which gives us:
\begin{equation}\label{equ: dichotomy }
	P^W_{1,2i+1}(u \le \upsilon |\sigma) = P^W_{1,i+1}(u \le \upsilon|\sigma)\, P^W_{1,i+1}(u \le \upsilon + \varsigma(i) |\sigma)
\end{equation}

Note that since the numbering of the IID draws is arbitrary, this can be extended to the existence of a function $\varsigma$ for permutation of $\setNsn{d(\sigma)}$.

For the LRA on RANDAO, this works with $\varsigma=1$ when splitting $\setNsn{2i}$ in the middle as previously. This comes from the fact that controlling an additional block at the end of an epoch doubles the number of possible seeds to choose from. The new ones result from having not proposed this additional block, and then having the same proposition configurations as before for the remaining blocks. That is why $d(\sigma)$ is a power of two and why $\varsigma=1$: the new possibilities have the same costs as before, plus 1 for the new discarded block.

\begin{theorem}
	Using the given notations, if a pMDP satisfies the condition from equation~\ref{equ: subintervals condition}, then we can compute the Bellman operator for the ex-ante states in $ \bigO{|\setSig|^2\, |\setR|\, \max d\, \log(\max d)} $. Additionally, if $\setU$ is smaller than the worst case ($|\setU|\in\bigO{|\setSig|\, |\setR|\, \max d}$), we have a complexity in $ \bigO{|\setU|\, |\setSig|\, \log(\max d)} $.
\end{theorem}

The pseudocode of the computation can be found in appendix~\ref{app: algorithms}, algorithm~\ref{alg:Efficient PW distribution pMDP}. \textbf{We end up with a significant improvement in the case of the LRA: the complexity has been lowered from the initial $\bigO{2^\kappa \kappa^{2^{\kappa +2}}} $ to a polynomial complexity $ \bigO{\kappa^4} $}. That is because we have $ |\setSig|=\kappa+1 $, but also $ \max d = 2^\kappa $, and $ |\setU| \in \bigO{\kappa^2}$ 

\section{Vanishing discount factor}
The previous ways to solve MDPs rely on the use of the discount factor $\gamma$ which adjusts the value of future rewards compared to that of the present rewards. In the case of the card game or the LRA, the discount factor can help represent the fact that the repetition of the game has an unknown end (a patch preventing the LRA for example). However, if one supposes that the repetition is infinite, we need to make a few changes. Indeed, in Ethereum an epoch lasts $6.4$ minutes, meaning that $\gamma$ would be extremely close to 1 and when $\gamma \rightarrow  1$, we also have the state values $V$ or $W$ diverging. Usually this means that the number of iterations to convergence will increase in $ \bigO{(1-\gamma)^{-1}} $. We refer the reader to~\cite{Solan_2022}, theorems 9.28 and 9.29, to see in details how this can be handled. In short, we do not look at the state values, but at the relative state values. That is, given a reference ex-ante state $\sigma_0$, we look at the relative value function $ \Delta W_{\sigma_0}^* (\sigma) := W^*(\sigma)-W^*(\sigma_0) $ which converges when $\gamma \rightarrow  1$ for a large class of MDPs in which ergodic\footnote{an MDP where all states are recurrent, no matter the policy} MDPs are part of, and our problems are ergodic. In practice, this allows to reach convergence in very few iterations (6 for the cards, 10 for LRA). 
The adapted Bellman operator is:
\begin{multline}\label{eq:NDBellManOperator}
	T^*\, \Delta W_{\sigma_0}(\sigma) = \esp\limits_{s\sim P(s|\sigma)} \left[ \max_{a\in\setNsn{d(\sigma)}} R(s,a ) + \gamma \Delta W_{\sigma_0}(s_a^\sigma) \right] \\ 
	- \esp\limits_{s\sim P(s|\sigma_0)} \left[ \max_{a\in\setNsn{d(\sigma_0)}} R(s,a) + \gamma \Delta W_{\sigma_0}(s_a^\sigma) \right] 
\end{multline}
And the adapted policy extraction:
\begin{equation}\label{eq:NDPolicyExtraction}
\pi^*(s) = a^*\in\argmax\limits_{a\in\setNsn{d(\sigma)}} \left[ s^R_a + \gamma \Delta W_{\sigma_0}^*(s^\sigma_a) \right]
\end{equation}

\section{Results}

The code for computing the following results is available at~\cite{AnonymousRepo}.

\subsection{Optimal ex-ante state value function $W^*$}\label{sec:resultW}

In Table~\ref{tab:resultValue}, one can see the relative ex-ante state values $\Delta W^*_{\sigma'}(\sigma) $ for $\gamma=1$ of our two problems (we stopped at $\sigma = 3$ for the LRA, but in~\cite{AnonymousRepo}, one can see the results for higher $\sigma$). Those tables enable the agent to compute the optimal strategy with equation~\ref{eq:NDPolicyExtraction}. For instance, we see that the club-diamond relative value is $ 2.01 $, this means that picking a 7 of diamond is worth slightly more than a 9 of club (without considering costs). Similarly, if revealing all blocks would yield $\sigma'=1$ (only 1 block controlled at the end of the next epoch) an attacker would be willing to lower his net reward by $1$ only if it allows him to get at least the last $3$ blocks instead (2 would be insufficient as the relative value $\sigma = 1$ to $\sigma = 2 $ is $0.97$, but the 1-3 relative value is $1.79$). 

\subsection{Runtimes for Variable Change Value Iteration on LRA}\label{sec:resultRuntime}

Table~\ref{tab-resultruntime} presents the runtimes of an iteration for the variable change value iteration of Section~\ref{sec:VariableChange}. They seem compatible with the theoretical complexity claim of an iteration complexity in  $\bigO{\kappa^4}$. The row $\log_2(\mathtt{ratio\ with\ previous\ time})$ should be exactly equal to 4 if the runtimes were exactly proportional to $\kappa^4$. Given the overhead involved in our algorithm this it makes perfect sense to have a lower ratio for lower problem sizes. The overheads have complexities in $\bigO{\kappa^2}$  and $\bigO{\kappa^3}$ respectively for the computation of the utility set $\setU$ and for the computation of the single-draw-distribution $P^W_{1,2}(u|\sigma)$ before computing the dichotomy Equation~\ref{equ: dichotomy }.

\begin{table*}[t]
	\caption{Average iteration runtime for the Variable change Value iteration for different LRA sizes}
	\label{tab-resultruntime}
	\centering
	\begin{tabular}{lrrrrrrrr}
		\toprule
		$\kappa$ & $8$ & $16$ & $32$ & $64$ & $128$ & $256$ & $512$ & $1024$\\
		\midrule
		time per iteration (ms) & $1.12\mathrm{e}{-1}$ & $2.68\mathrm{e}{-1} $ & $1.19\mathrm{e}{0} $ & $9.21\mathrm{e}{0}$ & $8.87\mathrm{e}{1}$ & $8.39\mathrm{e}{2}$ & $8.01\mathrm{e}{3}$ & $8.59\mathrm{e}{4}$\\
		$\log_2(\mathtt{ratio\ with\ previous\ time})$ & & $1.26$ & $2.15$ & $2.95$ & $3.27$ & $3.24$ & $ 3.26 $& $  3.42 $ \\
		\bottomrule
	\end{tabular}
	\caption*{Obtained with CPU: Intel Core Ultra 9 185H × 22}
\end{table*}

\subsection{Attack Performance}\label{sec:resultAttackPerf}

The obtained results in terms of attack performance are unimpressive but in line with the previous results of~\cite{OptimalRandaoManip}, which confirms the validity of our methodology. As they did, we compare the optimal attack to:
\begin{itemize}
	\item The myopic strategy: the attacker simply maximize the instant reward, it is myopic in the sense that it is reached for $\gamma = 0$, i.e. no value is given to future rewards.
	\item The control-max strategy: the attacker maximize the number of blocks controlled at the end  of the next epoch and ignores rewards.
	\item The honest strategy: no blocks are withheld.
\end{itemize}

\begin{figure}[ht]
	\begin{tikzpicture}
		\begin{axis}[
			xlabel={Attacker normalized stake},
			ylabel={Normalized additional average reward (\%)},
			title={Comparing strategies},
			legend pos=north west,
			grid=major,
			xmin=0, xmax=0.4,
			ymin=-0.01, ymax=0.04,
			]
			
			\addplot[color=blue] coordinates {
				(0.01,0.000011) (0.02,0.000062) (0.03,0.000161) (0.04,0.000303) (0.05,0.000483) (0.060000000000000005,0.000700) (0.06999999999999999,0.000949) (0.08,0.001230) (0.09,0.001541) (0.09999999999999999,0.001881) (0.11,0.002249) (0.12,0.002645) (0.13,0.003069) (0.14,0.003519) (0.15000000000000002,0.003996) (0.16,0.004499) (0.17,0.005029) (0.18000000000000002,0.005585) (0.19,0.006168) (0.2,0.006777) (0.21000000000000002,0.007413) (0.22,0.008077) (0.23,0.008768) (0.24000000000000002,0.009490) (0.25,0.010247) (0.26,0.011034) (0.27,0.011852) (0.28,0.012706) (0.29000000000000004,0.013593) (0.3,0.014516) (0.31,0.015476) (0.32,0.016476) (0.33,0.017518) (0.34,0.018603) (0.35000000000000003,0.019735) (0.36000000000000004,0.020917) (0.37,0.022154) (0.38,0.023450) (0.39,0.024810) (0.4,0.026244) 
			};
			\addlegendentry{Optimal strategy}
			
			\addplot[color=red] coordinates {
				(0.01,0.000011) (0.02,0.000062) (0.03,0.000160) (0.04,0.000299) (0.05,0.000477) (0.060000000000000005,0.000690) (0.06999999999999999,0.000935) (0.08,0.001210) (0.09,0.001515) (0.09999999999999999,0.001846) (0.11,0.002205) (0.12,0.002590) (0.13,0.003000) (0.14,0.003434) (0.15000000000000002,0.003893) (0.16,0.004376) (0.17,0.004882) (0.18000000000000002,0.005412) (0.19,0.005964) (0.2,0.006539) (0.21000000000000002,0.007137) (0.22,0.007757) (0.23,0.008399) (0.24000000000000002,0.009063) (0.25,0.009750) (0.26,0.010459) (0.27,0.011189) (0.28,0.011942) (0.29000000000000004,0.012717) (0.3,0.013514) (0.31,0.014333) (0.32,0.015175) (0.33,0.016039) (0.34,0.016927) (0.35000000000000003,0.017838) (0.36000000000000004,0.018775) (0.37,0.019744) (0.38,0.020756) (0.39,0.021836) (0.4,0.023042) 
			};
			\addlegendentry{Myopic strategy}
			
			\addplot[color=green!60!black] coordinates {
				(0.01,-0.000000) (0.02,-0.000000) (0.03,-0.000001) (0.04,-0.000002) (0.05,-0.000004) (0.060000000000000005,-0.000007) (0.06999999999999999,-0.000012) (0.08,-0.000018) (0.09,-0.000025) (0.09999999999999999,-0.000035) (0.11,-0.000046) (0.12,-0.000060) (0.13,-0.000076) (0.14,-0.000095) (0.15000000000000002,-0.000117) (0.16,-0.000141) (0.17,-0.000169) (0.18000000000000002,-0.000200) (0.19,-0.000234) (0.2,-0.000272) (0.21000000000000002,-0.000315) (0.22,-0.000361) (0.23,-0.000412) (0.24000000000000002,-0.000468) (0.25,-0.000529) (0.26,-0.000595) (0.27,-0.000667) (0.28,-0.000746) (0.29000000000000004,-0.000831) (0.3,-0.000924) (0.31,-0.001024) (0.32,-0.001133) (0.33,-0.001252) (0.34,-0.001381) (0.35000000000000003,-0.001520) (0.36000000000000004,-0.001672) (0.37,-0.001838) (0.38,-0.002019) (0.39,-0.002217) (0.4,-0.002434) 
			};
			\addlegendentry{Control-max strategy}

			\addplot[color=black] coordinates {
				(0.01,0.00) (0.40,0.00) 
			};
			\addlegendentry{Baseline - Honest strategy}
			
		\end{axis}
	\end{tikzpicture}
	\caption{Normalized additional average reward (percentage) as a function of relative stake for different strategies, $\kappa=32$}
	\label{fig-pMDPresultcurve}
	\Description{Curves of the Normalized additional average reward in percentage depending on the used strategy. even with large stakes, the gains are marginal (about 1.5\% for 30\% stake). Worse, they are barely better than the myopic strategy where agents simply select actions with the best momentary reward.}
\end{figure}
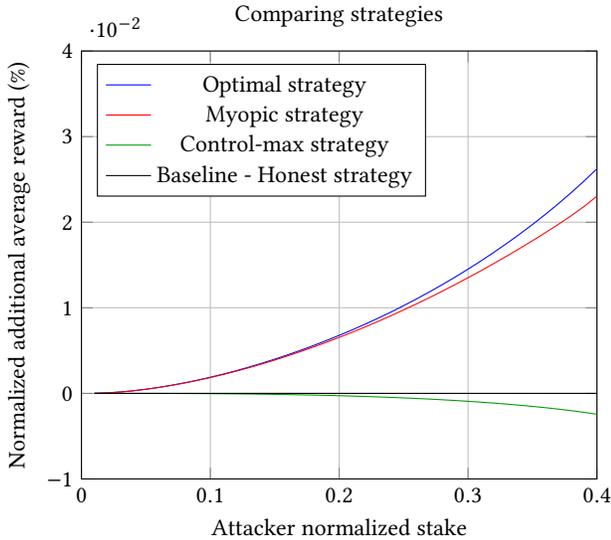
Figure~\ref{fig-pMDPresultcurve} shows that the optimal strategy offers only marginal gains over the myopic policy, and both deliver weak profitability. Hence a rational attacker would likely prefer the simpler, less bug-prone myopic approach; the optimal attack also bears higher financial risk by sometimes accepting lower current rewards. Considering additional deterrents, e.g., social slashing (community-enforced reputational or economic penalties), these attacks appear unattractive. The control-max policy then serves as a proxy for multi-attacker dynamics and, given its modest cost, could be a viable defensive strategy to limit others’ control (indeed, if you control the last block, you guarantee the other agents cannot attack).

\balance

Keep in mind however that this results stems from the strong hypotheses that all blocks are valued the same. As mentioned in~\cite{Randaomanip}, some blocks are much more rewarding than others, which would further reward attacks. But this would need to be modeled slightly differently with stochastic block rewards.

\section{Conclusion}
We introduced Pseudo Markov Decision Processes (pMDPs) as a framework to model and optimize stochastic decision problems with a reversed decision flow such as the Last Revealer Attack (LRA) in Ethereum’s RANDAO which was our case study. We reduced pMDPs to standard MDPs in two ways, then by leveraging the dual perspective of those reductions, we derived theoretical results enabling optimal policy computation and deeper insight into attacker strategies. We also identified conditions under which these methods apply to broader MDP subclasses, including settings where the discount factor plays a limited role—enhancing the applicability of pMDPs. This allowed us to reduce value iteration complexity per iteration from the initial $\bigO{2^\kappa \kappa^{2^{\kappa +2}}} $ to a polynomial complexity $ \bigO{\kappa^4} $, and to have it converge in just a few iterations. 

Empirically, our results align with this picture on three dimensions. \emph{Value computation.} We computed optimal ex-ante state values \(W^*\) (Section~\ref{sec:resultW}), which directly yield optimal policies via Equation~\ref{eq:NDPolicyExtraction} and quantify trade-offs between immediate net reward and end-of-epoch control. \emph{Runtime.} The variable-change value iteration shows iteration times consistent with the \(\bigO{\kappa^{4}}\) claim (Table~\ref{tab-resultruntime}) once overheads are taken into account. \emph{Strategy comparisons.} As reported in Figure~\ref{fig-pMDPresultcurve}, the optimal policy offers only marginal gains over the myopic policy (\(\gamma=0\)), and both yield weak profitability relative to the honest baseline. A rational attacker would therefore likely adopt the simpler myopic strategy, avoiding the additional implementation and financial risk of an “optimal” attack that occasionally sacrifices current rewards. Considering extra-protocol penalties (social slashing), such attacks are further disincentivized. We also evaluated a control-max policy as a proxy for multi-attacker dynamics; its cost is modest and it can act as a viable defensive heuristic (e.g., keeping last-block control to deny seed manipulation to others).

Overall, pMDPs provide a practical and interpretable tool for large, combinatorial decision problems where future control matters. Beyond the LRA, the same methodology applies to MDP subclasses with similar information structure such as the taxi ride selection problem presented in introduction. Limitations include sensitivity to cost function specifications, the limited size of the reversed flow decision problem class. Promising directions include: scaling via function approximation or learning; incorporating heterogeneous agents and explicit multi-agent interactions across epochs as well as stress-testing defensive policies (e.g., control-max variants) in adversarial simulations. We hope the combination of theory, runtime evidence, and strategy comparisons provides a useful basis for security assessment and mechanism design in proof-of-stake systems and related settings.

\bibliographystyle{ACM-Reference-Format} 
\bibliography{bibli.bib}

\newpage
\section{Proofs}
\subsection{Theorem~\ref{th: XA MDP}: Extracting optimal policy from ex-ante state values}\label{app:proofTHMPolicyDerivation}
\begin{proof}
	We first need to define the second MDP, and we do so similarly as in Section~\ref{sec:reduce}: with $ \setAsig = \{ \alpha: s\in \setS_\sigma \mapsto a\in\setAs$, this time we have $P(\sigma'|\sigma,\alpha) = \sum\limits_{s\in\setS} P(s|\sigma) P(\sigma'|s,\alpha(s))$, which is indeed a probability distribution thanks to the conditional independence hypothesis. And finally the reward function is $ R(\sigma,\alpha) = \esp(R(s,\alpha(s))|\sigma) $. With this, the tuple $(\setSig,\setAlph,P,R)$ is an MDP. 
	
	Then we need to prove that we can indeed substitute the expected value of $W^*$ in the optimal policy evaluation equation. To do that we first look at how the optimal value function $W^*$ relates to $V^*$, using the conditional independence and the Markov property, we have for $ \sigma'\in\setSig $:
	\begin{align*}
		W^*(\sigma') &= \esp\left[\sum\limits_{t=0}^\infty \gamma^t R_t| \sigma_{0}=\sigma',\pi^*\right]\\
		&=\sum\limits_{s'\in\setS} P(s'|\sigma',\pi^*) \esp\left[\sum\limits_{t=0}^\infty \gamma^t R_t| s_{0}=s',\sigma_{0}=\sigma',\pi^*\right]\\
		&=\sum\limits_{s'\in\setS} P(s'|\sigma') \esp\left[\sum\limits_{t=0}^\infty \gamma^t R_t| s_{0}=s',\pi^*\right]\\
		&= \esp\left[V^*(s')|\sigma'\right]
	\end{align*}
	
	Then using once again the conditional independence of $s'$ to $s,a$ when knowing $\sigma'$ and the previous equality, we re-write the expectation term in the optimal policy extraction:
	
	\begin{align*}
		\esp\left[V^*(s')|s,a\right] &= \sum\limits_{\sigma'\in\setSig} P(\sigma'|s,a) \esp\left[ V^*(s')|s,a,\sigma' \right]\\
		&= \sum\limits_{\sigma'\in\setSig} P(\sigma'|s,a) \esp\left[ V^*(s')|\sigma' \right]\\
		&=\sum\limits_{\sigma'\in\setSig} P(\sigma'|s,a) W^*(\sigma')\\
		&=\esp\left[W^*(\sigma')|s,a\right]
	\end{align*}
	
	This means that the policy extraction:
	\[ 
	\pi^* = s \in\setS \mapsto a^*\in\argmax\limits_{a\in\setAs} \left[ s^R_a + \gamma \esp\left(V^*(s')|s,a\right) \right]
	\]

	can indeed be re-written:
	\[ 
	\pi^* = s \in\setS \mapsto a^*\in\argmax\limits_{a\in\setAs} \left[ s^R_a + \gamma \esp\left(W^*(\sigma')|s,a\right) \right]
	\]
\end{proof}
\newpage 
\subsection{Bellman operator for ex-ante reduced pMDPs}\label{app:BellmanOperatorExAntepMDP}

Given a pMDP  $(\setSig,\setR,P,d,c)$, and its reduced MDPs $(\setS,\setA,P,R)$ and  $(\setSig,\setAlph,P,\mathrm{R})$
we first show that the Bellman operator for the ex-ante reduction satisfies equation~\ref{eq:BellmanExAnteMDP}. This satisfies the second result of theorem~\ref{thm: exAnte MDP value iteration}. 

\begin{proof}
	\begin{align*}
		T^*W(\sigma) & = \max_{\alpha \in\setAlph(\sigma)} \left[ R(\sigma,\alpha)+ \gamma \esp(W(\sigma')|\sigma,\alpha) \right]\\
		&= \max_{\alpha \in\setAlph(\sigma)} \left[ \esp (R(s,\alpha(s))|\sigma)+ \gamma \sum\limits_{\sigma'\in\setSig} P(\sigma'|\sigma,\alpha)\,W(\sigma') \right]\\
		&= \max_{\alpha \in\setAlph(\sigma)} \left[ \sum\limits_{s\in\setS} \left[P(s|\sigma) R(s,\alpha(s))\right]+ \gamma \sum\limits_{\sigma'\in\setSig}  \sum\limits_{s\in\setS} \left[ P(\sigma',s|\sigma,\alpha)\,W(\sigma')  \right]\right]\\
		&= \max_{\alpha \in\setAlph(\sigma)} \left[ \sum\limits_{s\in\setS} \left[P(s|\sigma) R(s,\alpha(s))\right]+ \gamma \sum\limits_{\sigma'\in\setSig}  \sum\limits_{s\in\setS} \left[ P(s|\sigma)P(\sigma'|s,\alpha(s))\,W(\sigma')  \right]\right]\\
		&= \max_{\alpha \in\setAlph(\sigma)} \left[ \sum\limits_{s\in\setS} P(s|\sigma) \left[ R(s,\alpha(s))+ \gamma \sum\limits_{\sigma'\in\setSig} P(\sigma'|s,\alpha(s)) \, W(\sigma')  \right]\right]\\
		&= \sum\limits_{s\in\setS} P(s|\sigma) \max_{\alpha \in\setAlph(\sigma)} \left[ R(s,\alpha(s))+ \gamma \sum\limits_{\sigma'\in\setSig} P(\sigma'|s,\alpha(s)) \, W(\sigma')  \right]\\
		&= \sum\limits_{s\in\setS} P(s|\sigma) \max_{a\in\setAs} \left[ R(s,a)+ \gamma \sum\limits_{\sigma'\in\setSig} P(\sigma'|s,a) \, W(\sigma')  \right]\\
		&= \esp_{s\sim P(s|\sigma)}\left[ \max_{a\in\setAs} \left[ R(s,a)+ \gamma \sum\limits_{\sigma'\in\setSig} P(\sigma'|s,a) \, W(\sigma')  \right]\right]\\
		&= \esp_{s\sim P(s|\sigma)}\left[ \max_{a\in\setAs} \left[ R(s,a)+ \gamma \sum\limits_{\sigma'\in\setSig} \ind{\sigma' = s_a^\sigma} \, W(\sigma')  \right]\right]\\
		&= \esp_{s\sim P(s|\sigma)}\left[ \max_{a\in\setAs} \left[ R(s,a)+ \gamma  W(s_a^\sigma)  \right]\right]
	\end{align*}
	
	The first part of the theorem is explicit in the last two lines of calculations: value iteration solely consist in repeatedly computing the Bellman operator, which will need iterations for each $\sigma\in\setSig,\, s\in\setS,\,a\in\setAs$; which is why an iteration complexity is in $ \bigO{|\setSig||\setS||\setA|} $. The pseudocode for this adapted value iteration can be found in appendix~\ref{app: algorithms}, algorithm~\ref{alg:ValueIterationExAnteReducedpMDP}.
	
\end{proof}
\newpage 
\subsection{Variable change}\label{app:variableChange}

Here we prove that the variable change of section~\ref{sec:VariableChange} allows the computation of the Bellman operator as the expected utility. The proof solely consist in the calculations showing exactly the targeted equation: $ \esp_{u\sim P^W(u|\sigma)} (u|\sigma)  = T^*W(\sigma) $. We place ourselves under the assumptions of theorem~\ref{th: XA MDP} and use the same notations as in the proof~\ref{app:proofTHMPolicyDerivation}

\begin{proof}
	
	For the calculations we use the utility probabilities as defined $P^W(u|\sigma) = \sum_{s\in\setS}P(s|\sigma) \ind{u^*_W(s) = u}$ (those are indeed probabilities for any $\sigma$ as they are positive and obviously add up to 1 when summed over $u\in\setU$). 
	
	\begin{align}
		\esp_{u\sim P^W(u|\sigma)} (u|\sigma) & = \sum\limits_{u \in \setU} P^W(u|\sigma) u\\
		& =  \sum\limits_{u \in \setU} \sum\limits_{s\in\setS}P(s|\sigma)\ind{u^*_W(s) = u} u\\
		& = \sum\limits_{s\in\setS}P(s|\sigma) \sum\limits_{u \in \setU} \ind{u^*_W(s) = u} u\\
		& = \sum\limits_{s\in\setS}P(s|\sigma) \, u^*_W(s) \\
		& = \sum\limits_{s\in\setS}P(s|\sigma) \, \max\limits_{a\in\setAs} \left[ R(s,a) + \gamma \esp (W(\sigma')|s,a) \right]\\
		& = \sum\limits_{s\in\setS}P(s|\sigma) \, \max\limits_{\alpha \in \setAsig} \left[ R(s,\alpha(s)) + \gamma \esp (W(\sigma')|s,\alpha(s)) \right]\\
		& = \max\limits_{\alpha \in \setAsig} \sum\limits_{s\in\setS}P(s|\sigma) \,  \left[ R(s,\alpha(s)) + \gamma \esp (W(\sigma')|s,\alpha(s)) \right]\\
		& = \max\limits_{\alpha \in \setAsig}  \left[ \esp (R(s,\alpha(s))|\sigma) + \gamma \esp (W(\sigma')|\sigma,\alpha) \right]\\
		& = \max\limits_{\alpha \in \setAsig}  \left[ R(\sigma,\alpha) + \gamma \esp (W(\sigma')|\sigma,\alpha) \right]\\
		& = T^*W(\sigma)
	\end{align}
\end{proof}

\section{Algorithms}\label{app: algorithms}
In this appendix, one can find the pseudocode for all algorithms mentioned in the paper.
We repeatedly use the fact that for a pMDP $(\setSig,\setR,P,d,c) $, with a reduced ex-post state set $\setS$ and reduced action set $\setA$ we have $\setS\in\bigO{(|\setSig|\, |\setR|)^{\max d}}$ and $|\setA| = \max d$.

\newcommand{\COMM}[1]{\hfill \textcolor{darkgray}{\textit{#1}}}

\begin{algorithm}[H]
	\caption{Standard value iteration}
	\label{alg:ValueIteration}
	\begin{algorithmic}
		\STATE {\bfseries Input:} MDP $ (\setS,\setA,P,R) $, discount factor $\gamma$, convergence precision $\varepsilon$
		\STATE Initialize $ \mathbf{V} := 0$, $ \mathbf{V'} := 0$ 
		\COMM{Vectorized versions of the value function}
		\REPEAT
		\STATE $ \mathbf{V} := \mathbf{V'}$
		\STATE $ \mathbf{V'} := -\infty$
		\FOR{$ s \in \setS $\COMM{computing $ \mathbf{V'} = T^*\mathbf{V}$}}
		\FOR{$ a \in \setAs $\COMM{computing one element $ \mathbf{V'}[s] = \max_{a\in\setAs} \left[R(s,a) + \gamma \esp(V(s')|s,a)\right] $}}
		\STATE Initialize $v_{s,a}:=R(s,a)$
		\FOR{$ s' \in\setS $ \COMM{computing the term in $ \gamma \esp(V(s')|s,a) $}}
		\STATE $ v_{s,a} \pluseq \gamma P(s'|s,a)\, \mathbf{V}[s'] $
		\ENDFOR
		\STATE $ \mathbf{V'}[s] := \max (\mathbf{V'}[s],v_{s,a}) $
		\ENDFOR
		\ENDFOR
		\UNTIL{$||\mathbf{V}-\mathbf{V}'|| \le \varepsilon$}
		\STATE {\bfseries Output:} Value vector $\mathbf{V}$
	\end{algorithmic}
\end{algorithm}
Algorithm~\ref{alg:ValueIteration}: Loops on $ \setS $, $ \setAs $, $ \setS $. Complexity in $ \bigO{|\setS|^2\,|\setA|} $

\begin{algorithm}[H]
	\caption{Value iteration for reduced pMDPs}
	\label{alg:ValueIterationExAnteReducedpMDP}
	\begin{algorithmic}
		\STATE {\bfseries Input:} pMDP $ (\setSig,\setR,P,d,c) $, ex-post state set $\setS$, distributions $P(s|\sigma)$, discount factor $\gamma$, convergence precision $\varepsilon$
		\STATE Initialize $ \mathbf{W} := 0$, $ \mathbf{W'} := 0$ \COMM{Vectorized versions of the ex-ante value function}
		\REPEAT
		\STATE $ \mathbf{W} := \mathbf{W'}$
		\STATE $ \mathbf{W'} := 0$
		\FOR{$\sigma \in \setSig$ \COMM{computing $ \mathbf{W'} = T^*\mathbf{W} $, using equation~\ref{eq:BellmanExAnteMDP}} }
		\FOR{$ s \in \setS $ \COMM{computing one element $ \mathbf{W'}[\sigma] = \esp_{s\sim P(s|\sigma)} \left[ \max_{a\in\setAs} R(s,a ) + \gamma W(s_a^\sigma) \right]$}}
		\STATE $ V := -\infty $
		\FOR{$ a \in \setNsn{d(\sigma)} $ \COMM{computing a possible outcome $  \max_{a\in\setAs} R(s,a ) + \gamma W(s_a^\sigma)$}}
		\STATE $ {V} := \max ({V}[s], s^r_a - c(a) + \gamma \mathbf{W}[s_a^\sigma]) $ \COMM{Note that for reduced pMDPs $ R(s,a) = s_a^r -c(a) $}
		\ENDFOR
		\STATE $ \mathbf{W'}[\sigma] \pluseq  P(s|\sigma)\, V $
		\ENDFOR
		\ENDFOR
		\UNTIL{$||\mathbf{W}-\textbf{W}'|| \le \varepsilon$}
		\STATE {\bfseries Output:} Value vector $\mathbf{W}$
	\end{algorithmic}
\end{algorithm}

Algorithm~\ref{alg:ValueIterationExAnteReducedpMDP}: Loops on $ \setSig $, $ \setS $, $ \setA = \setNsn{d(\sigma)} $. Complexity in $ \bigO{|\setSig|\,|\setS|\,|\setA|} \subset  \bigO{|\setSig|(|\setSig|\, |\setR|)^{\max d}\, \max d} $

\begin{algorithm}[H]
	\caption{Value iteration for MDPs of theorem~\ref{th: XA MDP}}
	\label{alg:ValueIterationMDPwithCond}
	\begin{algorithmic}
		\STATE {\bfseries Input:} MDP $ (\setS,\setA,P,R) $, set $\setSig$, distributions $ P(\sigma'|s,a) $ and $ P(s|\sigma) $, discount factor $\gamma$, convergence precision $\varepsilon$
		\STATE Initialize $ \mathbf{W} := 0$, $ \mathbf{W'} := 0$ \COMM{Vectorized versions of the 'ex-ante' value function}
		\REPEAT
		\STATE $ \mathbf{W} := \mathbf{W'}$
		\STATE $ \mathbf{W'} := 0$
		\FOR{$\sigma \in \setSig$ \COMM{computing $ \mathbf{W'} = T^*\mathbf{W} $, using equation~\ref{eq:BellmanExAnteMDP}}}
		\FOR{$ s \in \setS $ \COMM{computing one element $ \mathbf{W'}[\sigma] = \esp_{s\sim P(s|\sigma)} \left[ \max_{a\in\setAs} R(s,a ) + \gamma\, \esp_{\sigma'\sim P(\sigma'|s,a)} \left[W(\sigma') \right]\right]$}}
		\STATE $ V := -\infty $
		\FOR{$ a \in \setAs $\COMM{computing a possible outcome $ V:= \max_{a\in\setAs} R(s,a ) + \gamma\, \esp_{\sigma'\sim P(\sigma'|s,a)} \left[W(\sigma') \right]$}}
		\STATE $ V' := R(s,a)$\COMM{computing $ V':=R(s,a ) + \gamma\, \esp_{\sigma'\sim P(\sigma'|s,a)} \left[W(\sigma') \right]$}
		\FOR{$ \sigma' \in \setSig $\COMM{adding the term in $ \gamma\, \esp_{\sigma'\sim P(\sigma'|s,a)} \left[W(\sigma') \right]$}}
		\STATE $ V' \pluseq \gamma\,P(\sigma'|s,a) \, \mathbf{W}[\sigma']$
		\ENDFOR
		\STATE $ {V} := \max ({V}, V') $
		\ENDFOR
		\STATE $ \mathbf{W'}[\sigma] \pluseq P(s|\sigma) \,V$
		\ENDFOR
		\ENDFOR
		\UNTIL{$||\mathbf{W}-\mathbf{W}'|| \le \varepsilon$}
		\STATE {\bfseries Output:} Value vector $\mathbf{W}$
	\end{algorithmic}
\end{algorithm}

Algorithm~\ref{alg:ValueIterationMDPwithCond}: Loops on $ \setSig $, $ \setS $, $ \setA $, $ \setSig $. Complexity in $ \bigO{|\setSig|^2\,|\setS|\,|\setA|} $

\begin{algorithm}[H]
	\caption{Standard Policy extraction for a single state}
	\label{alg:Policy Derivation}
	\begin{algorithmic}
		\STATE {\bfseries Input:} MDP $ (\setS,\setA,P,R) $, discount factor $\gamma$, Value vector $ \mathbf{V} $, state $s\in\setS$
		\STATE Initialize $ V' := -\infty$
		\FOR{$ a \in \setAs $ \COMM{computing $ a^*\in\argmax_{a\in\setAs} R(s,a) + \gamma\, \esp_{s'\sim P(s'|s,a)} \left[V(s')\right] $}}
		\STATE Initialize $v_{s,a}: = R(s,a)$
		\FOR{$ s' \in\setS $\COMM{adding the term $  \gamma\, \esp_{s'\sim P(s'|s,a)} \left[V(s')\right] $}}
		\STATE $ v_{s,a} \pluseq \gamma P(s'|s,a)\, \mathbf{V}[s'] $
		\ENDFOR
		\IF{$v_{s,a}>V'$}
		\STATE $ a^* := a $
		\STATE $ V' := v_{s,a} $
		\ENDIF
		\ENDFOR
		\STATE {\bfseries Output:} action $a^*$
	\end{algorithmic}
\end{algorithm}

Algorithm~\ref{alg:Policy Derivation}: Loops on $ \setA $, $ \setS $. Complexity in $ \bigO{|\setS|\,|\setA|} $

\begin{algorithm}[H]
	\caption{Policy extraction for ex-ante reduced pMDP, for a single ex-post state}
	\label{alg:Policy Derivation exAnte reduced pMDP}
	\begin{algorithmic}
		\STATE {\bfseries Input:} pMDP $ (\setSig,\setR,P,d,c) $, discount factor $\gamma$, Value vector $ \mathbf{W} $, state $s\in\setS$
		\STATE Initialize $ W' := -\infty$
		\FOR{$ a \in \setAs = \setNsn{d(s)} $ \COMM{computing $ a^*\in\argmax_{a\in\setAs} s^r_a-c(a) + \gamma\, W(s^\sigma_a) $}}
		\STATE $ w_{s,a} := s^r_a-c(a) + \gamma\, \mathbf{W}[s^\sigma_a] $
		\IF{$w_{s,a}>W'$}
		\STATE $ a_{best} := a $
		\STATE $ W' := w_{s,a} $
		\ENDIF
		\ENDFOR
		\STATE {\bfseries Output:} action $a_{best}$
	\end{algorithmic}
\end{algorithm}

Algorithm~\ref{alg:Policy Derivation exAnte reduced pMDP}: Loops on $ \setA $. Complexity in $ \bigO{|\setA|} \subset  \bigO{\max d} $

\begin{algorithm}[H]
	\caption{Policy extraction for MDPs of theorem~\ref{th: XA MDP} for a single state}
	\label{alg:Policy Derivation exAnte}
	\begin{algorithmic}
		\STATE {\bfseries Input:} MDP $ (\setS,\setA,P,R) $, set $\setSig$, distributions $ P(\sigma'|s,a) $, discount factor $\gamma$, Value vector $ \mathbf{W} $, state $s\in\setS$
		\STATE Initialize $ W' := -\infty$
		\FOR{$ a \in \setAs $\COMM{computing $ a^*\in\argmax_{a\in\setAs} R(s,a) + \gamma\, \esp_{\sigma'\sim P(\sigma'|s,a)} \left[W(\sigma')\right] $}}
		\STATE Initialize $w_{s,a} := R(s,a)$
		\FOR{$ \sigma' \in\setSig $\COMM{adding the term $ \gamma\, \esp_{\sigma'\sim P(\sigma'|s,a)} \left[W(\sigma')\right] $}}
		\STATE $ w_{s,a} \pluseq \gamma P(\sigma'|s,a)\, \mathbf{W}[\sigma'] $
		\ENDFOR
		\IF{$w_{s,a}>W'$}
		\STATE $ a_{best} := a $
		\STATE $ W' := w_{s,a} $
		\ENDIF
		\ENDFOR
		\STATE {\bfseries Output:} action $a_{best}$
	\end{algorithmic}
\end{algorithm}

Algorithm~\ref{alg:Policy Derivation exAnte}: Loops on $ \setA $, $ \setSig $. Complexity in $ \bigO{|\setSig|\,|\setA|} $

\begin{algorithm}[H]
	\caption{Monte Carlo Value Iteration for MDPs of theorem~\ref{th: XA MDP} }
	\label{alg:MCVI MDP from thm}
	\begin{algorithmic}
		\STATE {\bfseries Input:} MDP $ (\setS,\setA,P,R) $, set $\setSig$, distributions $ P(\sigma'|s,a) $ and $ P(s|\sigma) $, discount factor $\gamma$, convergence precision $\varepsilon$, sample size $ n_{sample}$
		\STATE Initialize $ \mathbf{W} := 0$ and $ \mathbf{W'} := 0$ \COMM{Vectors of the size of  $ \setSig $}
		\REPEAT
		\STATE $ \mathbf{W} := \mathbf{W'}$ \COMM{Decreasing learning rates can be used to force convergence: $ \mathbf{W} \pluseq rate( \mathbf{W'}-  \mathbf{W} )  $}
		\STATE $ \mathbf{W'} := 0$
		\FOR{$\sigma \in \setSig$ \COMM{computing $ \mathbf{W'} = T^*\mathbf{W} $, using equation~\ref{eq:BellmanExAnteMDP}}}
		\FOR{$ i \in \setNsn{n_{sample}}$ \COMM{MC evaluation of $ \mathbf{W'}[\sigma] = \esp_{s\sim P(s|\sigma)} \left[ \max_{a\in\setAs} R(s,a ) + \gamma\, \esp_{\sigma'\sim P(\sigma'|s,a)} \left[W(\sigma') \right]\right]$}}
		\STATE Sample $s\in\setS$ with probability $ P(s|\sigma) $ \COMM{For reduced pMDP, sampling $s = (s_i)_{i\in\setNsn{d(\sigma)}}$} is  in $ \bigO{d(\sigma)} $
		\FOR{$ a \in \setAs $\COMM{computing $ \max_{a\in\setAs} R(s,a) + \gamma\, \esp_{\sigma'\sim P(\sigma'|s,a)} \left[W(\sigma')\right] $}}
		\STATE Initialize $w_{s,a} := R(s,a)$
		\FOR{$ \sigma' \in\setSig $\COMM{adding the term $ \gamma\, \esp_{\sigma'\sim P(\sigma'|s,a)} \left[W(\sigma')\right] $}}
		\STATE $ w_{s,a} \pluseq \gamma P(\sigma'|s,a)\, \mathbf{W}[\sigma'] $ \COMM{For reduced pMDPs, skip the loop with $ w_{s,a} = s_a^r - c(a) + \mathbf{W}[s_a^\sigma] $}
		\ENDFOR
		\STATE $  \mathbf{W}'[\sigma] \pluseq w_{s,a} $
		\ENDFOR
		\ENDFOR
		\STATE $ \mathbf{W}'[\sigma] :=\mathbf{W}'[\sigma] /n_{sample} $
		\ENDFOR
		\STATE  $ \mathbf{W} := \mathbf{W'}  $ 
		\UNTIL{$||W-W'|| \le \varepsilon$}
		\STATE {\bfseries Output:}  Value vector $\mathbf{W}$
	\end{algorithmic}
\end{algorithm}

Algorithm~\ref{alg:MCVI MDP from thm}: Loops on $ \setSig $, $ \setNsn{n_{sample}} $, $ \setA $, $ \setSig $. Therefore the complexity is in $ \bigO{|\setSig|^2\,|\setA|\,n_{sample}\, \varsigma} $ (where $\varsigma$ is the sampling cost). For reduced pMDP the innermost loop can be skipped since we can directly compute $ w_{s,a} = s_a^r - c(i) + \gamma\,\mathbf{W}[s^\sigma_a]$. This allows us to lower the complexity to $ \bigO{|\setSig|\,|\setA|\,n_{sample}\, \varsigma } \subset  \bigO{|\setSig|\, \max d\, n_{sample}\, \varsigma} $.

\begin{algorithm}[H]
	\caption{Naïve computation of $P^W(u|\sigma)$ for reduced pMDPs}
	\label{alg: Naive PW distribution}
	\begin{algorithmic}
		\STATE {\bfseries Input:} 
		pMDP $ (\setSig,\setR,P,d,c) $, Value vector $ \mathbf{W} $, discount factor $\gamma$\\
		\STATE Initialize $\setU=\{\}$
		\FOR{$ \sigma\in\setSig,\, r\in\setR,\, i\in\setNsn{d(\sigma)} $ \COMM{We first compute $ \setU $}}
		\STATE $ \setU $.add$( r-c(i)+\gamma \, W(\sigma) )$
		\ENDFOR\\
		\STATE Initialize $ \mathbf{P}^W := zeroes(|\setSig|,|\setU|)$ \COMM{Matrix of zeroes of size $ |\setSig|*|\setU| $}
		\FOR{$ \sigma \in \setSig $ \COMM{proceeding ex-ante state by ex-ante state}} 
		\FOR{$ s \in\setS $ \COMM{Computing $ P^W[u |\sigma] = \sum_{s\in\setS} P(s|a) \ind{\max_{a\in\setNsn{d(\sigma)}} s^r_a-c(a)+\gamma\, \mathbf{W}[s^\sigma_a] = u}$}}
		\STATE Initialize $u:=-\infty$
		\FOR{$ a \in \setNsn{d(\sigma)} = \setAs $ \COMM{Computing $ \max_{a\in\setNsn{d(\sigma)}} s^r_a-c(a)+\gamma\, \mathbf{W}[s^\sigma_a]$}}
		\STATE $ u:=\max(u,s^r_a-c(i)+\gamma \mathbf{W}[s^\sigma_a]) $
		\ENDFOR
		\STATE $ \mathbf{P}^W [\sigma,u] \pluseq P(s|\sigma) $ 
		\ENDFOR
		\ENDFOR
		\STATE {\bfseries Output:} Utility distribution $\mathbf{P}^W$
	\end{algorithmic}
\end{algorithm}

Algorithm~\ref{alg: Naive PW distribution}: Loops on $ \setSig $, $ \setS$, $ \setA $. Complexity in $ \bigO{|\setSig|\,|\setS|\,|\setA|} \subset  \bigO{|\setSig|(|\setSig|\, |\setR|)^{\max d}\, \max d} $. The initial computation of $ \setU $ is in $\bigO{|\setU|} \subset \bigO{|\setSig|\,|\setR|\,\max d}$ as utilities $u$ are additions of three terms. One is of the form $\gamma\,W(\sigma)$ for a $\sigma\in\setSig$, one is of the form $r\in\setR$, and one of the form $c(a)$ for $a\in\setA=\setNsn{\max_{d(\setSig)}} $. Note that $c(\setA)$ can be much smaller than $A$, for example if $c=0$, which is why $\bigO{|\setU|} \subset \bigO{|\setSig|\,|\setR|\,|\setA|}$; Additionally the term in $r-c(a)$ can sometimes be restricted to a small set (for instance in our card game, $r-c(a)$ is an integer between $ -13 $ and $ 3 $). In the LRA $r-c(a) \in [-32,32]$, so all in all the computation of $\setU$ is in $\bigO{\kappa^2}$.

\begin{algorithm}[H]
	\caption{Efficient computation of $P^W(u\le\upsilon|\sigma)$ for pMDPs}
	\label{alg:Efficient PW distribution pMDP}
	\begin{algorithmic}
		\STATE {\bfseries Input:} 
		pMDP $ (\setSig,\setR,P,d,c) $, Value vector $ \mathbf{W} $, discount factor $\gamma$, Probability adaptation function$ \varsigma $\\
		
		\STATE Initialize $\setU=\{\}$
		\FOR{$ \sigma\in\setSig,\, r\in\setR,\, i\in\setNsn{d(\sigma)} $ \COMM{We first compute $ \setU $}}
		\STATE $ \setU $.add$( r-c(i)+\gamma \, W(\sigma) )$
		\ENDFOR
		\STATE Initialize $ \mathbf{P}^W = zeroes(|\setSig|,|\setU|)$\COMM{Then we compute the probabilities} 
		\FOR{$ \sigma \in \setSig $} 
		\FOR{$ \sigma'\in\setSig $ \COMM{We initialize $\mathbf{P}^W [\sigma,\ .\ ] $ as if $d(\sigma)=1$}}
		\FOR{$ r\in\setR $}
		\STATE $ u = r-c(1)+\gamma\, W(\sigma')$
		\STATE $\mathbf{P}^W [\sigma,u] \pluseq P(\sigma',r|\sigma) $
		\ENDFOR
		\ENDFOR
		\STATE $\ $ \COMM{The following loop computes  $\mathbf{P}^W [\sigma,\ .\ ] $ as if $d(\sigma)=2^i$}
		\FOR{$ i\in\setNsn{\log_2(d(\sigma))} $ \COMM{For simplicity we consider only the cases where $d(\sigma)$ is a power of 2}}
		\STATE $\mathbf{P}^W [\sigma,:] := \mathbf{P}^W [\sigma,:] * \mathbf{P}^W [\sigma,: (+ \varsigma(i))]$ \COMM{last indexation is the implementation of equation~\ref{equ: subintervals condition}}
		\ENDFOR
		\ENDFOR
		\STATE {\bfseries Output:} Utility cumulative distributions $\mathbf{P}^W$
	\end{algorithmic}
\end{algorithm}

Algorithm~\ref{alg:Efficient PW distribution pMDP}: First we have the initial computation of $\setU$, as explained for the previous algorithm, that is in $\bigO{|\setU|} \subset \bigO{|\setSig|\,|\setR|\,\max d}$. Then we have loops on $ \setSig $, $ \setSig $, $ \setR $, this part is difficultly the bottleneck (for instance in the LRA's case it is in $ \bigO{\kappa^3} $), so we will ignore it. Then there are loops on  $ \setSig $, $\setNsn{\log_2(d(\sigma))}$. This gives a complexity in $ \bigO{|\setSig|\, \log |\setA|\, |\setU|} \subset \bigO{|\setSig|\, \log (\max d)\, |\setU|} \subset  \bigO{|\setSig|^2\, |\setR|\, \max d\, \log (\max d)} $

\begin{algorithm}[H]
	\caption{Value iteration without discount factor}
	\label{alg: ValueIteration No Discount}
	\begin{algorithmic}
		\STATE {\bfseries Input:} MDP $ (\setS,\setA,P,R) $, convergence precision $\varepsilon$
		\STATE Initialize $ \mathbf{\Delta V} := 0$, $ \mathbf{\Delta V'} := 0$ 
		\COMM{Vectorized versions of the value function}
		\REPEAT
		\STATE $ \mathbf{\Delta V} := \mathbf{\Delta V'} $ 
		\STATE compute the Bellman operator $ \mathbf{\Delta V'} := T^*\mathbf{\Delta V}$
		\STATE Center back: $ \mathbf{\Delta V'} = \mathbf{\Delta V'} -\mathbf{\Delta V'}[0] $
		\UNTIL{$||\mathbf{\Delta V}-\mathbf{\Delta V}'|| \le \varepsilon$}
		\STATE {\bfseries Output:} Value vector $\mathbf{\Delta V}$ \COMM{ This is the vector of values compared to that of the reference state 0}
	\end{algorithmic}
\end{algorithm}

\end{document}